\documentclass[letters,fleqn,usenatbib]{mnras}
\usepackage{newtxtext,newtxmath,graphicx,amsmath,hyperref}
\usepackage[T1]{fontenc}
\usepackage[dvipsnames]{xcolor}
\DeclareRobustCommand{\VAN}[3]{#2}
\let\VANthebibliography\thebibliography
\def\thebibliography{\DeclareRobustCommand{\VAN}[3]{##3}\VANthebibliography}
\newcommand{\be}{\begin{equation}}
\newcommand{\ee}{\end{equation}}
\newcommand{\bq}{\begin{eqnarray}}
\newcommand{\eq}{\end{eqnarray}}
\title[CMB temperature cosmography]{Cosmic microwave background temperature cosmography}
\author[C. J. A. P. Martins]{
C. J. A. P. Martins$^{1,2}$\thanks{E-mail: Carlos.Martins@astro.up.pt}\\
$^{1}$Centro de Astrof\'{\i}sica da Universidade do Porto, Rua das Estrelas, 4150-762 Porto, Portugal\\
$^{2}$Instituto de Astrof\'{\i}sica e Ci\^encias do Espa\c co, CAUP, Rua das Estrelas, 4150-762 Porto, Portugal\\
}
\date{Accepted XXX. Received YYY; in original form ZZZ}
\pubyear{2026}
\begin{document}
\label{firstpage}
\pagerange{\pageref{firstpage}--\pageref{lastpage}}
\maketitle

\begin{abstract}
The temperature-redshift relation, $T(z)=T_0(1+z)$, is a cornerstone of the standard cosmological model, but it is violated in many physically motivated extensions thereof, e.g. when photon number is not conserved. We apply the methodology of cosmography, a model-independent approach to cosmology, to this observable, and use astrophysical Sunyaev-Zeldovich and high-resolution spectroscopic measurements spanning the redshift range $0\le z\le6.34$, to constrain violations of the standard relation. We explore the impact of different cosmographic approaches and expansion variable choices and also compare the cosmographic constraints with those obtained, with the same data, in the canonical adiabatic extension model of Lima. While these choices have some impact, overall we find that such violations are constrained to percent level, and that a single non-standard parameter suffices to describe them. We also show that the directly fitted slope of the temperature-redshift relation is compatible with (but slightly smaller than) the one inferred from FIRAS, and that standard cosmography lore, based on scale factor expansions, does not apply to this observable, e.g. Pad\'e approximants have no meaningful advantage over plain Taylor series.
\end{abstract}
\begin{keywords}
(Cosmology:) cosmic background radiation -- (Cosmology:) cosmological parameters -- Cosmology: theory -- Methods: statistical
\end{keywords}

%%%%%%%%%%%%%%%%%%%%%%%%%%%%%%%%%%%%%%%%%%%%%%%%%%%%%%%%%%%%%%%%%%%%%%%%%%
\section{Introduction}
\label{introd}

There is growing observational evidence that the cosmological concordance model is an approximation to a more fundamental one. Moreover, the interpretation of our cosmological observations is model-dependent, usually relying \textit{a priori} assumptions on an underlying cosmological model---which may be a physical model, but is often a mathematical toy model that is naively subjected to a statistical analysis. Ideally, one would like to test the key assumptions of our cosmological paradigm using methods that are as model-independent as possible. In what follows we discuss a simple example of this.

Cosmography is an approach \citep{Visser}, in which physical quantities are expanded as a Taylor series, either in the cosmological redshift $z$ or related variables, around redshift zero. This advantage is offset by possible convergence issues beyond $z=1$ (which can be mitigated by using other variables \citep{Cattoen}), by the need to decide at which order to truncate the series (a precision vs. accuracy trade-off) and by potential degeneracies between the series coefficients. There may also be some dependence on the choices of priors for these coefficients \citep{Dunsby}. Still, one can envisage high-redshift cosmography \citep{Vitagliano,Pade2,Bargiacchi}, and it has also been applied to the redshift drift of objects following the cosmological expansion \citep{Rocha} and to dimensionless fundamental couplings \citep{Alpha}. Here we apply it to the temperature of the cosmic microwave background (CMB). One of our results is that some of the standard cosmography lore, based on expansions of the scale factor, does not apply here.

In physical cosmology, if the expansion of the Universe is adiabatic and the CMB spectrum was initially a black-body, this property is preserved by the cosmological evolution, with its temperature evolving as
\be\label{stdT}
T(z) = T_0(1+z) \,,
\ee
where $T_0$ is the present-day value. However, many standard model extensions violate this, concomitantly with violating the distance duality relation \citep{Duality}. The best motivated scenarios involve photon number non-conservation, through photon mixing due to coupling with particles beyond the particle physics standard model \citep{Avgoustidis}, but common astrophysical processes can also be erroneously interpreted as non-standard physics. The canonical phenomenological description of these violations, introduced in \citet{Lima1} and \citet{Lima2}, is
\be\label{limaT}
T(z) = T_0(1+z)^{1-\beta} \,:
\ee
in what follows we will compare constraints on $\beta$ with those obtained from cosmographic expansions.

We carry out standard maximum likelihood analyses, with 63 measurements of the CMB temperature in the redshift range $0\le z\le6.34$, and reporting one-sigma constraints unless otherwise stated. At redshift zero we use the FIRAS measurement of \citet{Fixsen}, 
\be\label{t0fixsen}
T_{z=0}=2.7255\pm0.0006\, {\rm K}\,.
\ee
At non-zero redshift there are measurements from two methods. At lower redshifts we rely on the thermal Sunyaev-Zeldovich effect: specifically from 815 Planck clusters in 18 redshift bins \citep{Hurier}, 158 SPT clusters in 12 redshift bins, and 15 earlier measurements \citep{SZ1,SZ2}. At higher redshifts we use high-resolution spectroscopy, with 17 publicly available measurements \citep{Srianand,Ge,Molaro,Cui,Muller,Klimenko,Riechers,Kotani,Kotani2}. Their cosmological impact has been highlighted, for specific models, in \citet{Gelo}; here we instead adopt a cosmographic approach.

%%%%%%%%%%%%%%%%%%%%%%%%%%%%%%%%%%%%%%%%%%%%%%%%%%%%%%%%%%%%%%%%%%%%%%%%%%%%%%
\section{Standard cosmography}
\label{meth}

\begin{figure*}
\includegraphics[width=0.51\columnwidth]{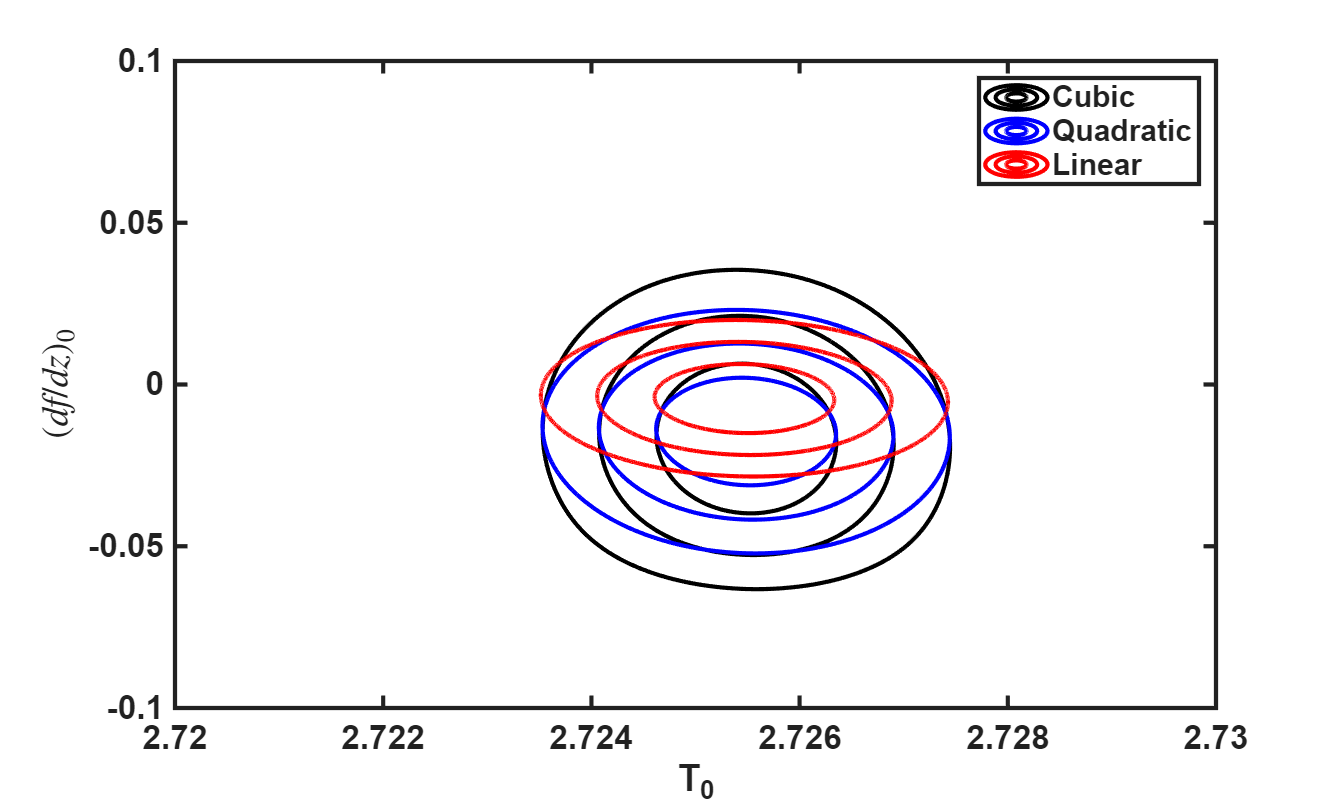}
\includegraphics[width=0.51\columnwidth]{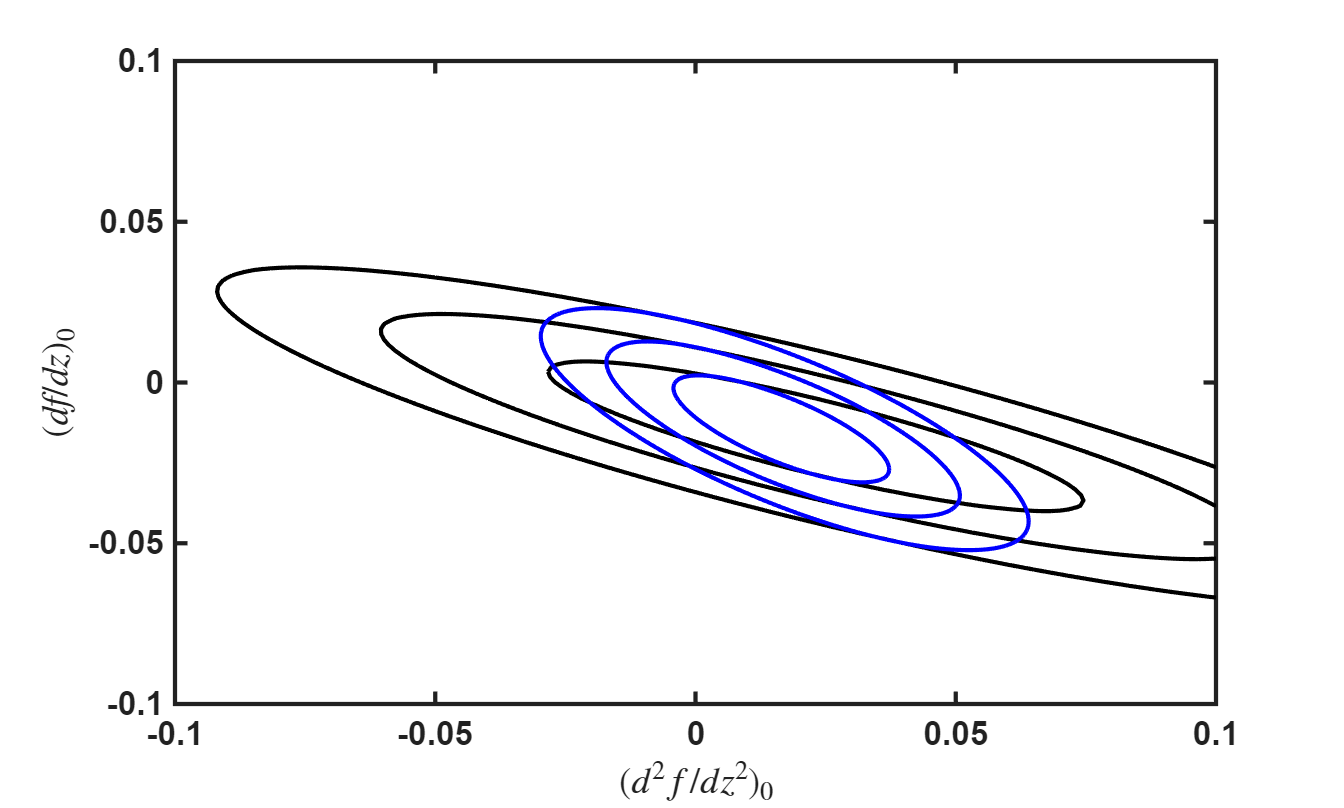}
\includegraphics[width=0.51\columnwidth]{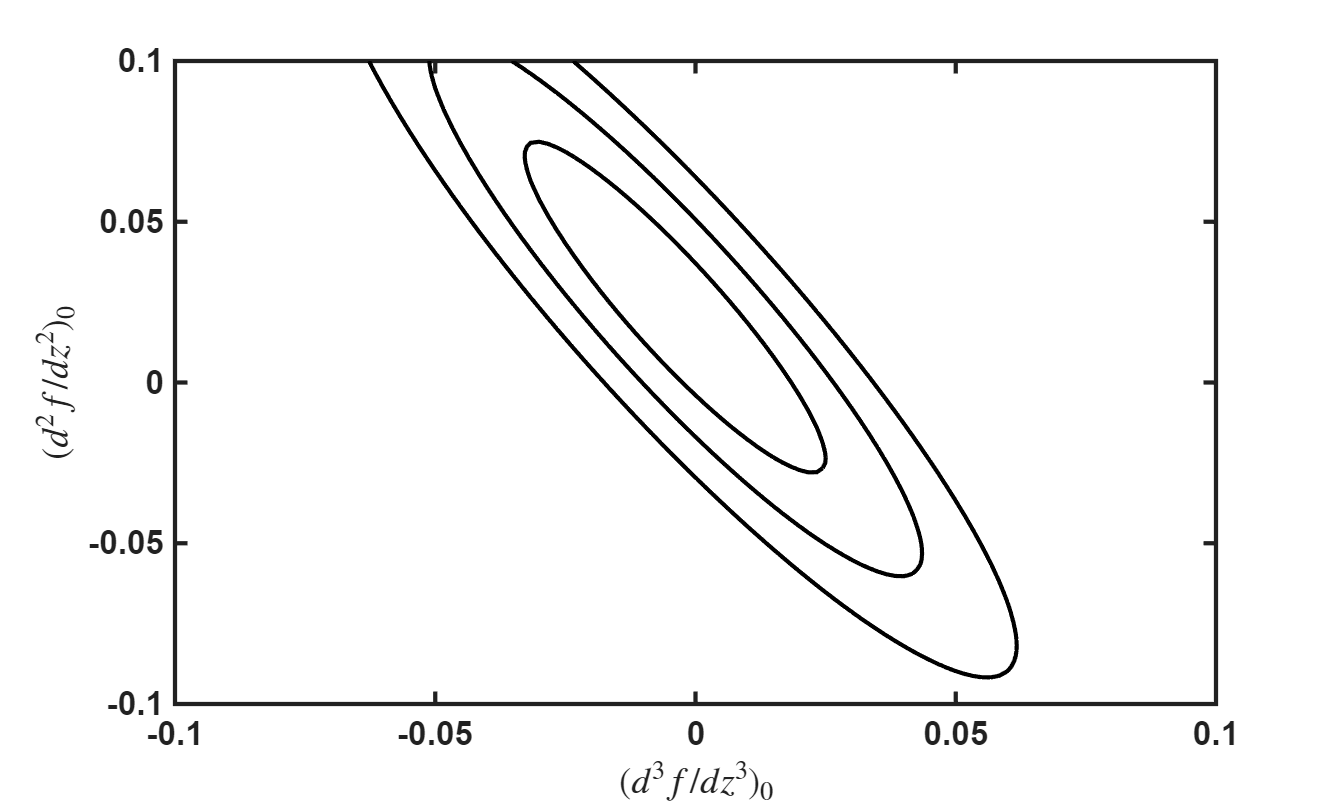}
\includegraphics[width=0.51\columnwidth]{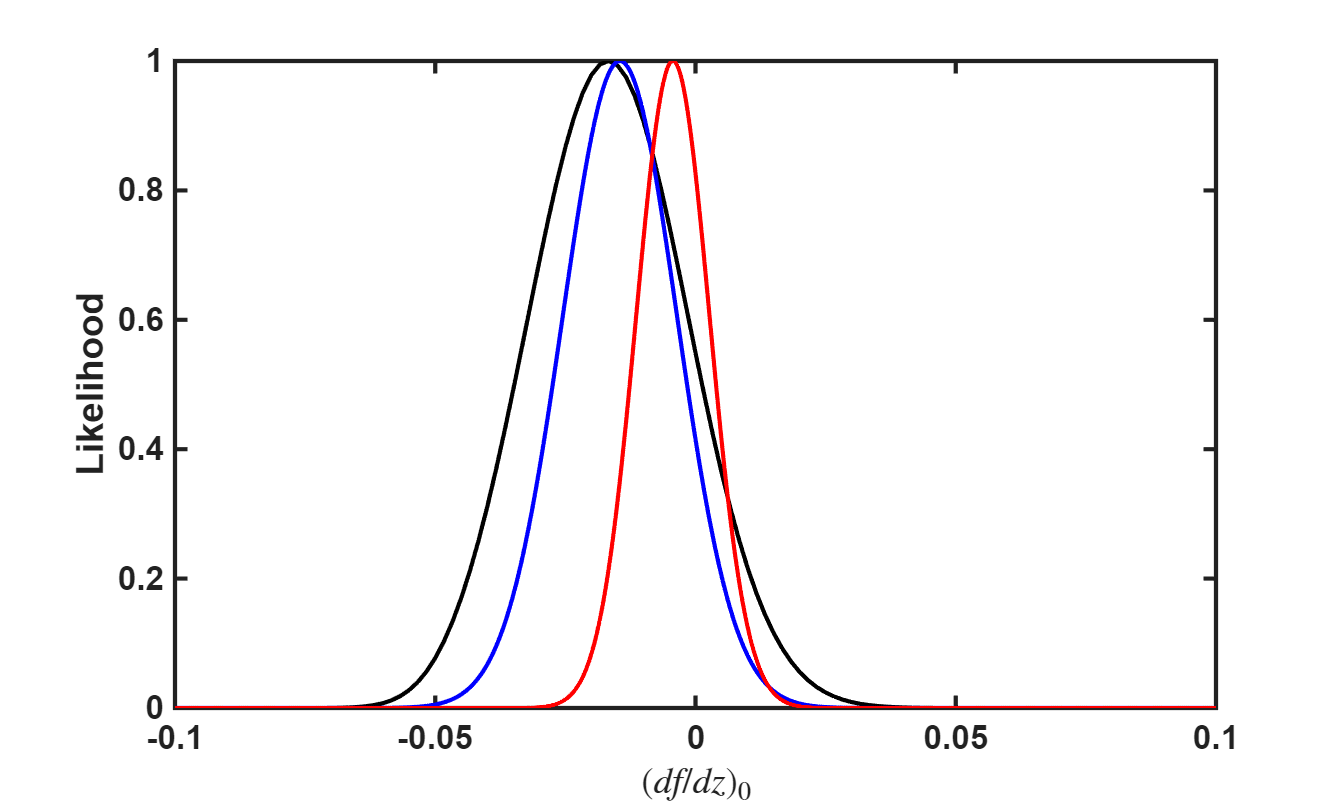}
\includegraphics[width=0.51\columnwidth]{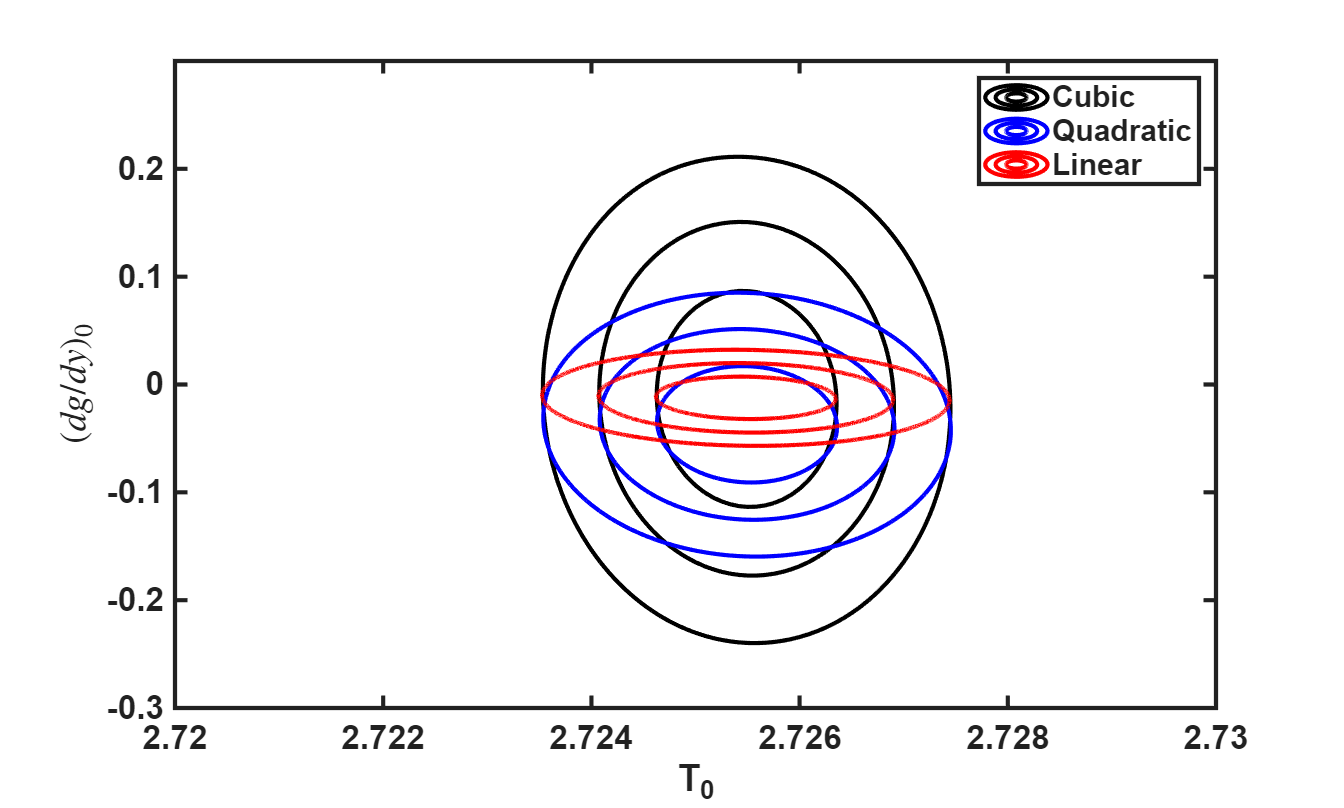}
\includegraphics[width=0.51\columnwidth]{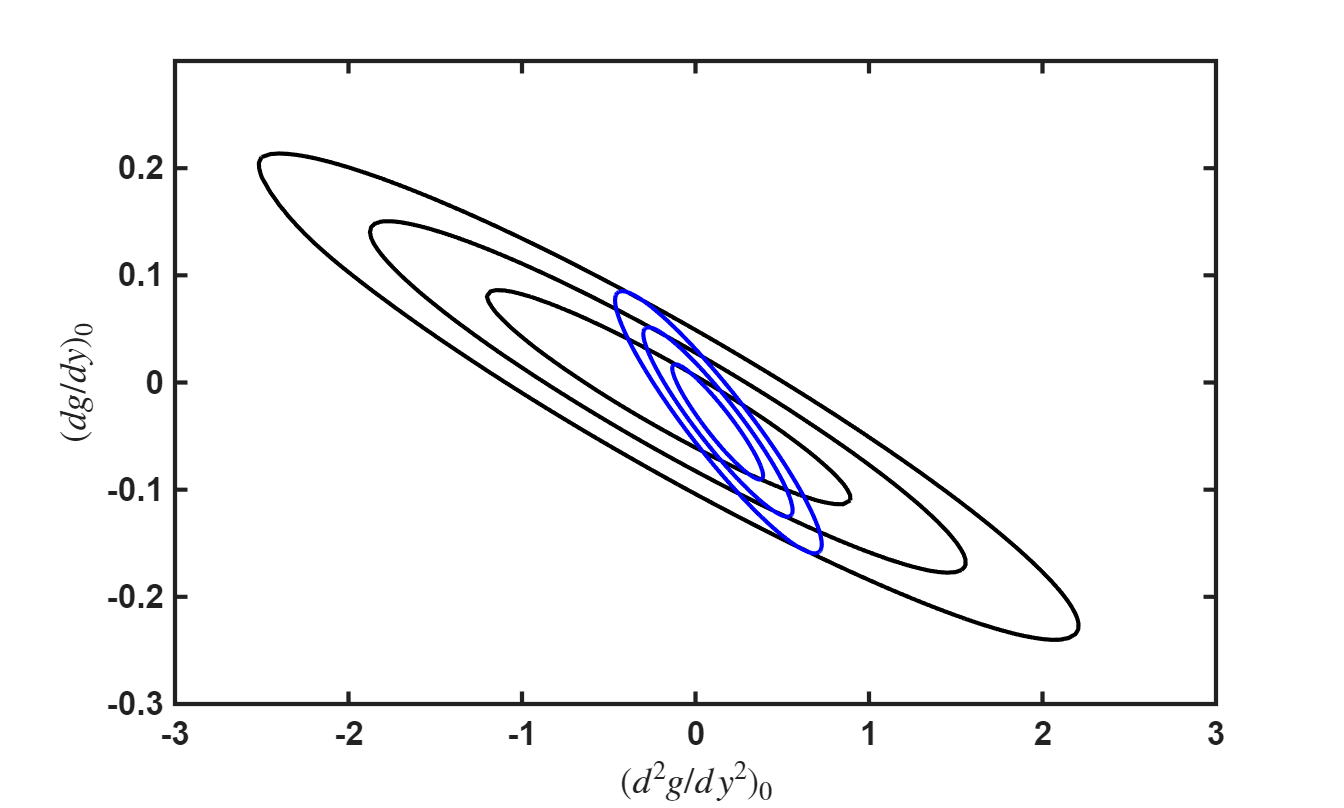}
\includegraphics[width=0.51\columnwidth]{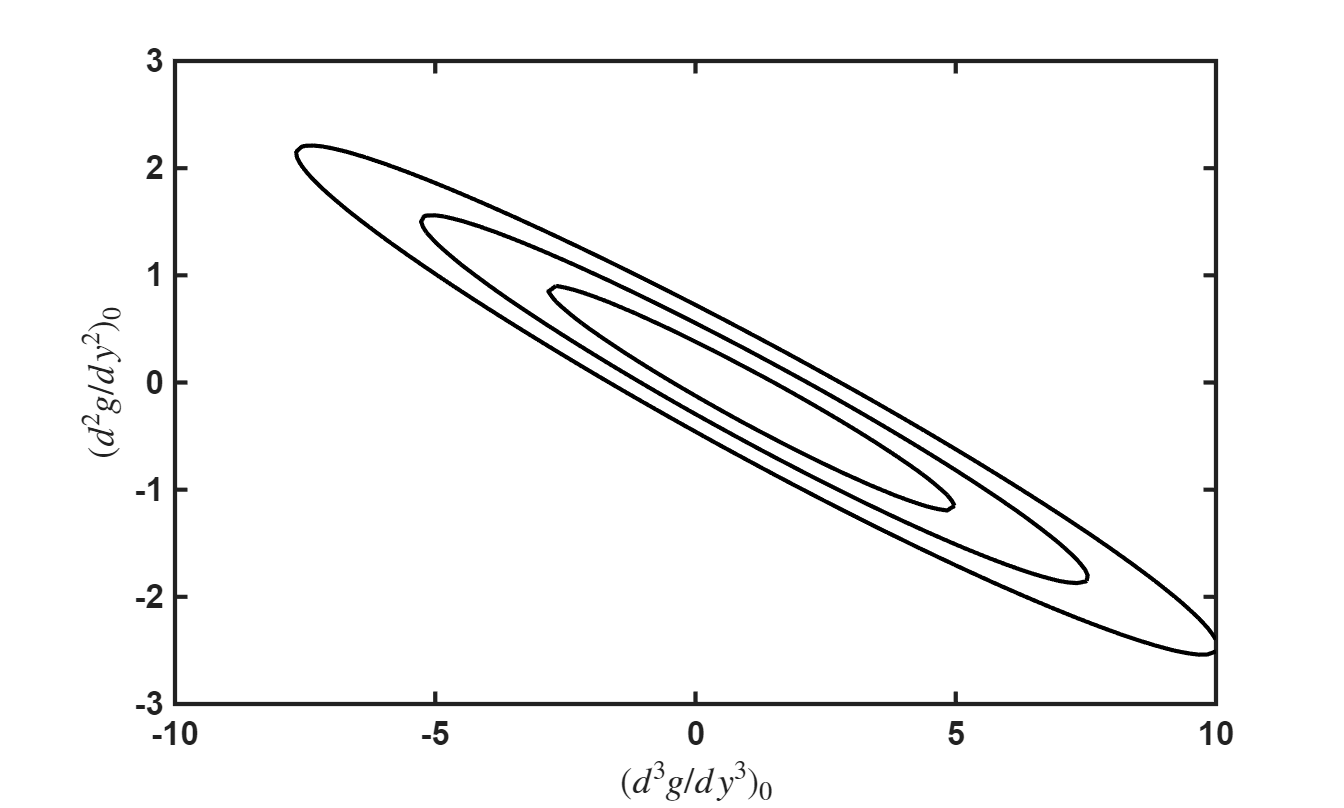}
\includegraphics[width=0.51\columnwidth]{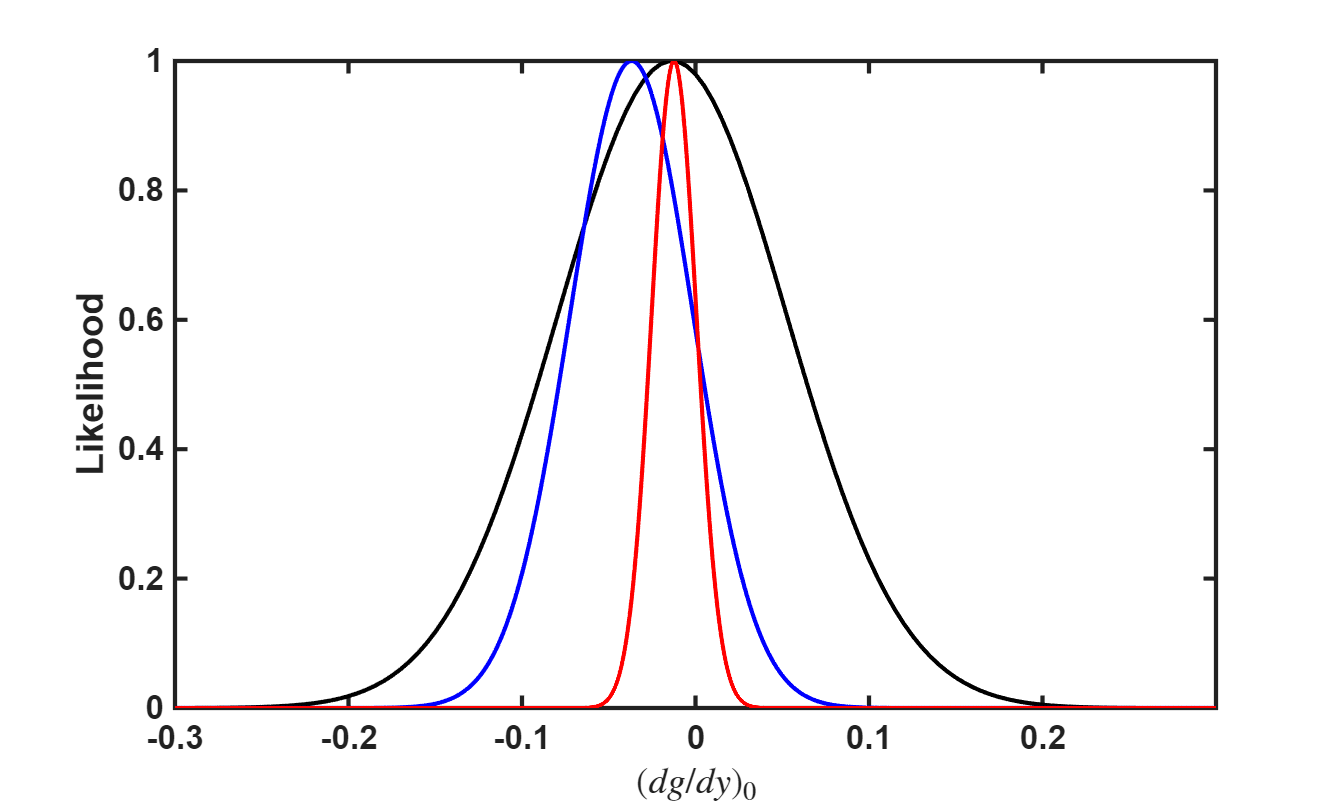}
\caption{Constraints on the parameters of the cosmographic series of Eq. (\ref{expansion0}) and Eq. (\ref{expansion1}), in the top and bottom panels respectively, as a function of the truncation order. In two-dimensional panels, one, two and three sigma confidence levels are shown. Note the different axis ranges in the top and bottom panels.}
\label{figure1}
\end{figure*}

Consider a generic redshift dependent correction to the standard law
\begin{equation}
T(z)=T_0(1+z)f(z)\,,
\end{equation}
\begin{equation}\label{expansion0}
f(z)=1+\left(\frac{df}{dz}\right)_0z+ \frac{1}{2}\left(\frac{d^2f}{dz^2}\right)_0z^2 + \frac{1}{6}\left(\frac{d^3f}{dz^3}\right)_0z^3+\cdots \,.
\end{equation}
A possible difficulty is series convergence, considering the availability of data at redshift $z>1$. One might therefore define a rescaled redshift $y=z/(1+z)$, and
\begin{equation}\label{expansion1}
g(y)=1+\left(\frac{dg}{dy}\right)_0y+ \frac{1}{2}\left(\frac{d^2g}{dy^2}\right)_0y^2 + \frac{1}{6}\left(\frac{d^3g}{dy^3}\right)_0y^3+\cdots \,.
\end{equation}
Table \ref{table1} and Figure \ref{figure1} summarize the resulting constraints on the parameters of these series, as a function of their truncation order. In all cases the constraint on $T_0$, which is also a free parameter, is given by the FIRAS measurement, and there is no evidence for corrections to the standard law. The two approaches are superficially comparable, although the coefficients in the $f(z)$ series are, naturally, constrained to be smaller, so superficially the $y$ expansion brings no meaningful gain. Contiguous coefficients, e.g. the first and second, and the second and third derivatives, are anticorrelated, as illustrated, for the former case, in the middle panel of Fig. \ref{figure1}.

%%%%%%%%%%%%%%%%%%%%
\begin{table}
\centering
\caption{One-sigma constraints on the parameters of the Eqs. (\ref{expansion0}--\ref{expansion1}) cosmographic series, as a function of the truncation order.}
\label{table1}
\begin{tabular}{c c c c}
\hline
Parameter & Linear & Quadratic & Cubic \\
\hline
$(df/dz)_0$ & $-0.004\pm0.007$ & $-0.014\pm0.011$ & $-0.017\pm0.016$ \\
$(d^2f/dz^2)_0$ & - & $0.016\pm0.013$ & $0.022\pm0.034$ \\
$(d^3f/dz^3)_0$ & - & - & $-0.003\pm0.019$ \\
\hline
$(dg/dy)_0$ & $-0.012\pm0.013$ & $-0.04\pm0.04$ & $-0.02\pm0.07$ \\
$(d^2g/dy^2)_0$ & - & $0.13\pm0.17$ & $-0.15\pm0.70$ \\
$(d^3g/dy^3)_0$ & - & - & $1.0\pm2.6$ \\
\hline
\end{tabular}
\end{table}
%%%%%%%%%%%%%%%%%%%%
%%%%%%%%%%%%%%%%%%%%

\begin{figure*}
\includegraphics[width=1.0\columnwidth]{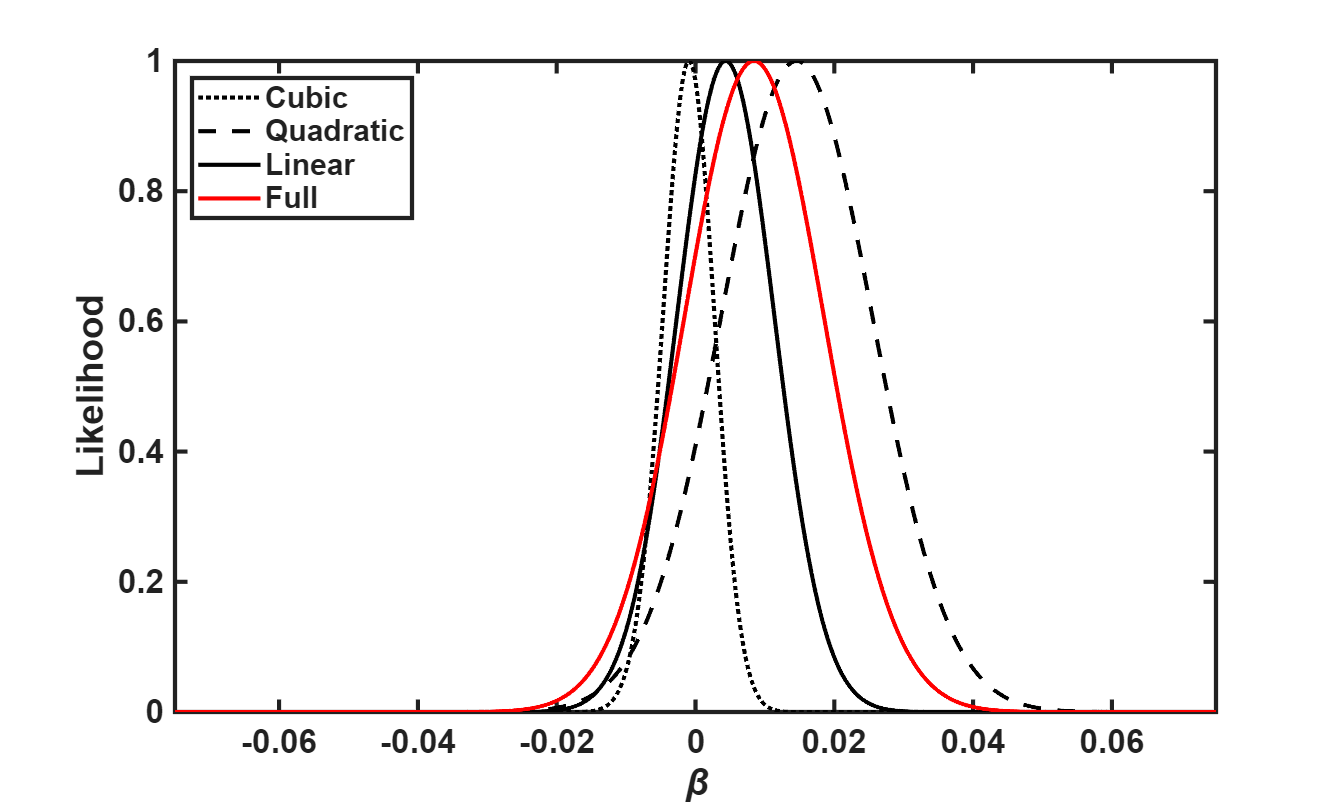}
\includegraphics[width=1.0\columnwidth]{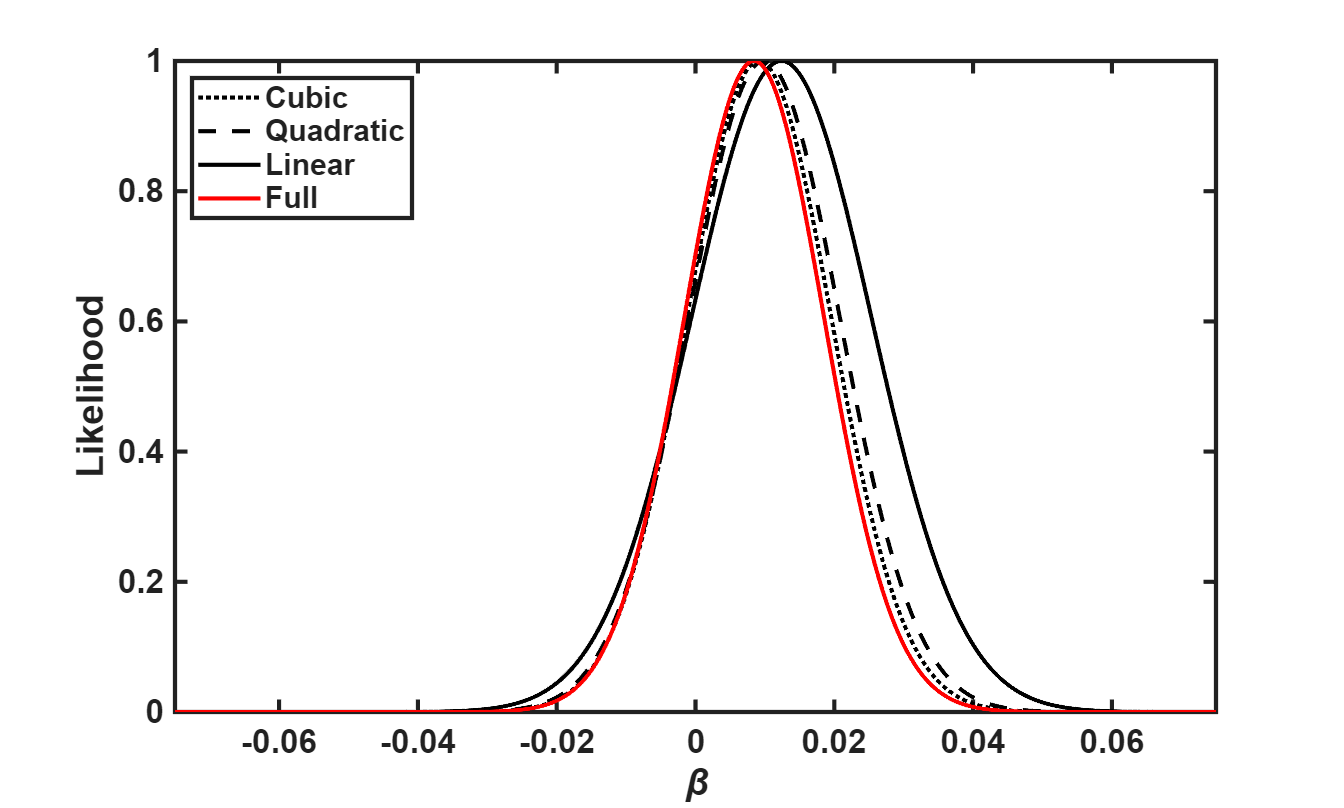}
\caption{Constraints on the Lima phenomenological temperature-redshift relation Eq. (\ref{limaT}) and its Taylor series approximation, truncated at various orders. Left and right panels correspond to the $f(z)$ and $g(y)$ series respectively.}
\label{figure2}
\end{figure*}

For the standard phenomenological description of Eq. (\ref{limaT}), we have $f(z)=(1+z)^{-\beta}$, or $g(y)=(1-y)^{\beta}$, for which
\be
\left(\frac{d^if}{dz^i}\right)_0=(-1)^i \prod_{j=1}^i(\beta+i-1)
\ee
\be
\left(\frac{d^ig}{dy^i}\right)_0=(-1)^i \prod_{j=1}^i(\beta-i+1)\,.
\ee
This is an opportunity to compare the constraints on $\beta$ from the Taylor series approximation, truncated at various orders, with the full power law function $f$. Table \ref{table2} and Figure \ref{figure2} show the corresponding results. Again as expected, the $T_0$ constraint is entirely dominated by FIRAS. Here the $f(z)$ series highlights the well-known convergence issues, while the $g(y)$ series is stable and provides a good approximation to the full behavior, slightly overestimating these uncertainties, especially in the linear case. This analysis also shows that a single beyond standard model physics suffices to characterise the data. In passing, we note that if we constrain the Lima model without the FIRAS measurement, we find
\bq
T_0&=&2.7246\pm0.0135\, {\rm K}\\
\beta&=&0.008\pm0.016\,,
\eq
with the two parameters being positively correlated.

%%%%%%%%%%%%%%%%%%%%
\begin{table}
\centering
\caption{One-sigma constraints on the Lima phenomenological temperature-redshift relation parameter $\beta$, defined in  Eq. (\ref{limaT}), and its Taylor series approximations, truncated at various orders. For the full case, the $95\%$ constraint is $\beta=0.008\pm0.020$.}
\label{table2}
\begin{tabular}{c c c}
\hline
Approximation & $f(z)$ series & $g(y)$ series\\
\hline
Linear & $0.004\pm0.007$ & $0.012\pm0.013$ \\
Quadratic & $0.015\pm0.011$ & $0.010\pm0.011$ \\
Cubic & $-0.001\pm0.004$ & $0.009\pm0.010$ \\
\hline
Full (Power law) & $0.008\pm0.010$ & $0.008\pm0.010$ \\
\hline
\end{tabular}
\end{table}
%%%%%%%%%%%%%%%%%%%%
%%%%%%%%%%%%%%%%%%%%

%%%%%%%%%%%%%%%%%%%%%%%%%%%%%%%%%%%%%%%%%%%%%%%%%%%%%%%%%%%%%%%
\section{Pad\'e cosmography}
\label{pade}

Pad\'e approximants \citep{Press} have been auggested as suitable for cosmographic descriptions \citep{Pade1}. Given a function $f(x)$, the Pad\'e approximant of order $[m/n]$, for $m\ge0$ and $n\ge1$, is the rational function
\be
P_{m,n}(x)=\frac{a_0+a_1x+a_2x^2+\cdot+a_mx^m}{1+b_1x+b_2x^2+\cdot+b_nx^n}\,,
\ee
whose coefficients are determined by the condition that values of the function and all its derivatives up to order $(m+n)$, at $x=0$, coincide with those with the function $f$ itself. As compared to polynomial functions of conventional Taylor series they are typically less oscillatory, can fit a wider range of curves and have well-behaved asymptotes. On the other hand, they are vulnerable to vertical nuisance asymptotes \citep{Guthrie}, and there is no clear prescription to choose the optimal degrees of the numerator and denominator; evidently, the choice depends on the function under consideration. For the scale factor and quantities derived from it, $P_{2,1}$ seems a good choice \citep{Pade2}, but the situation is known to be different for the redshift drift, as shown in \citet{Rocha}.

\begin{figure*}
\includegraphics[width=0.51\columnwidth]{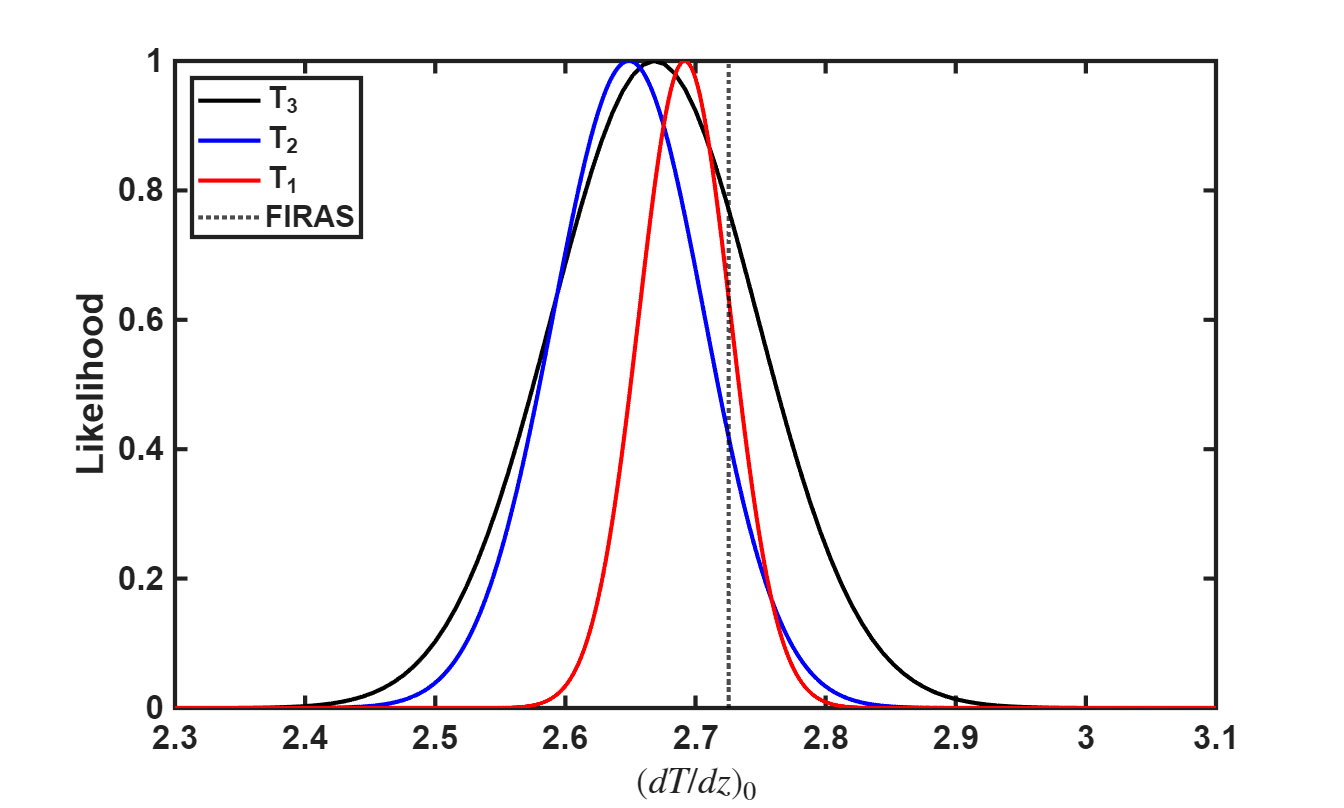}
\includegraphics[width=0.51\columnwidth]{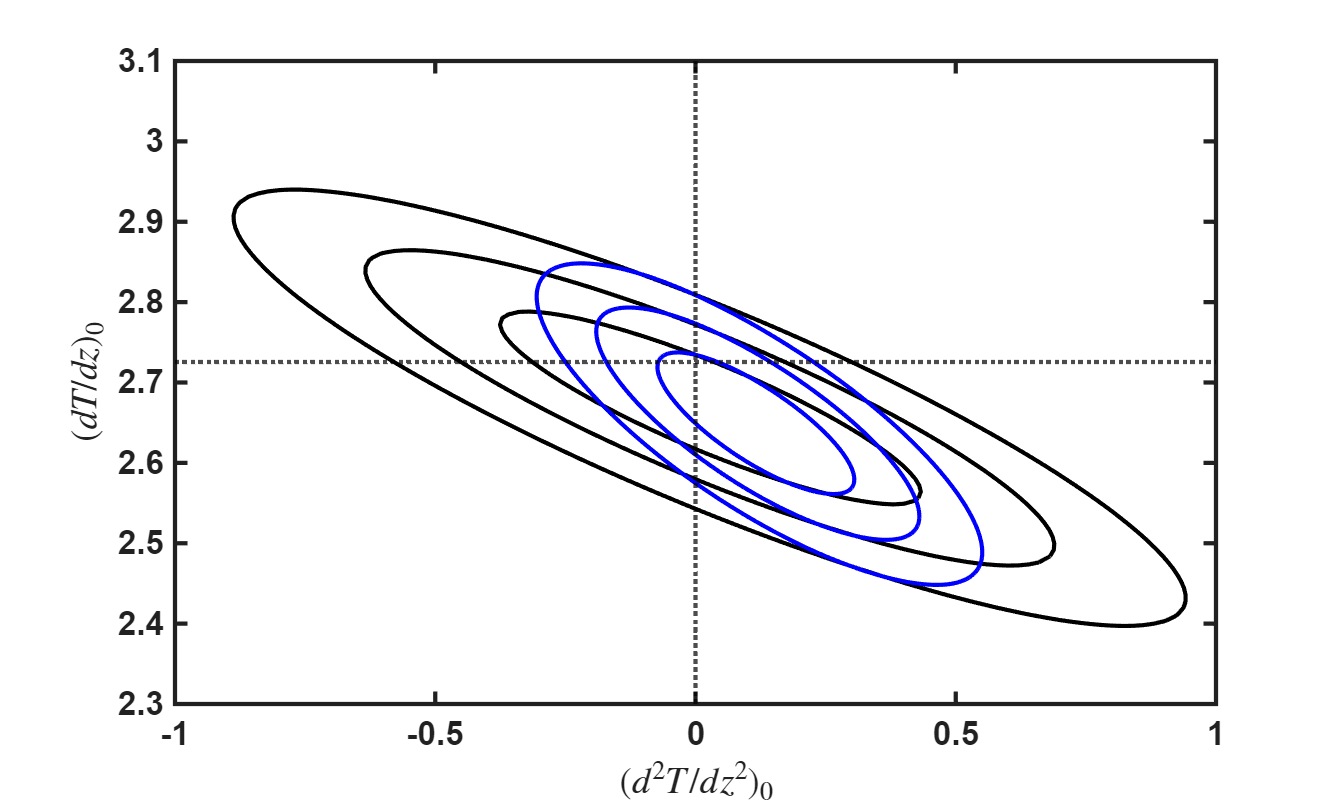}
\includegraphics[width=0.51\columnwidth]{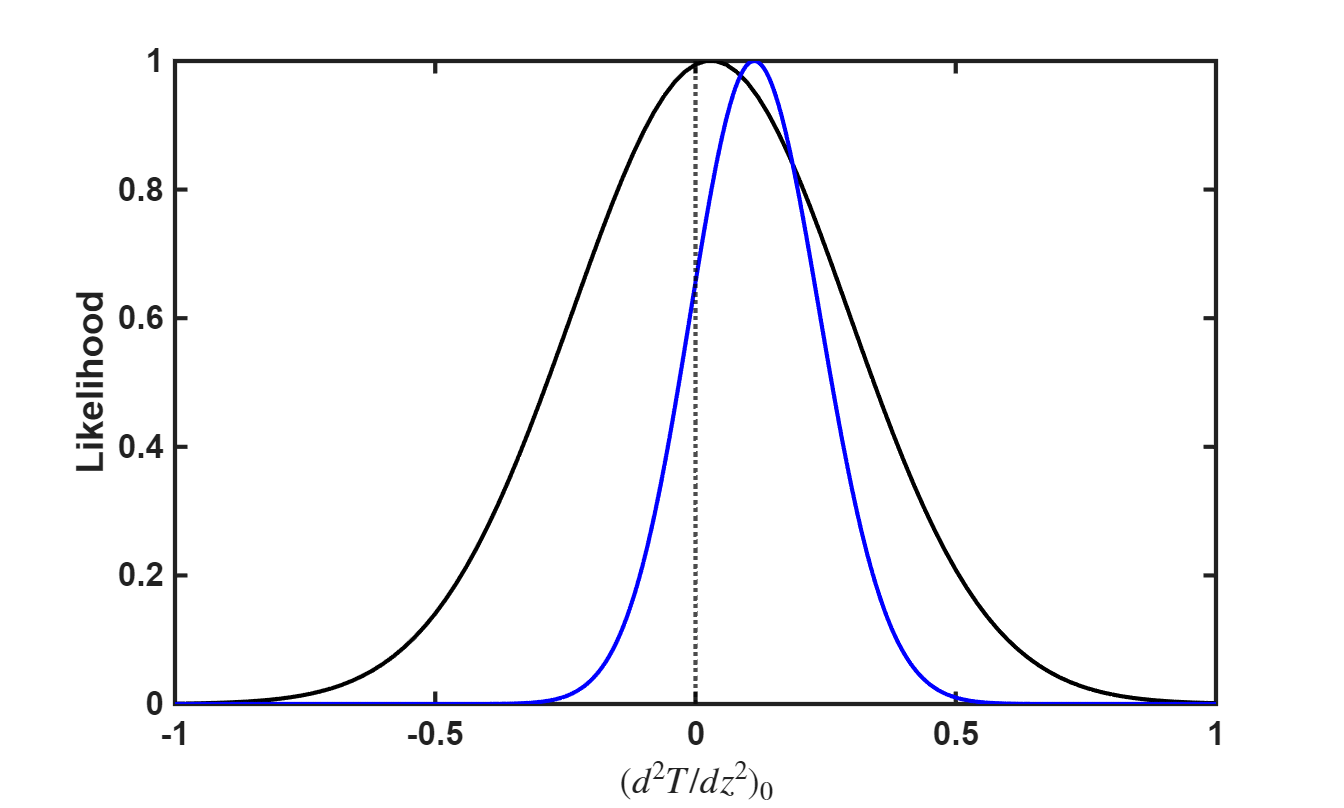}
\includegraphics[width=0.51\columnwidth]{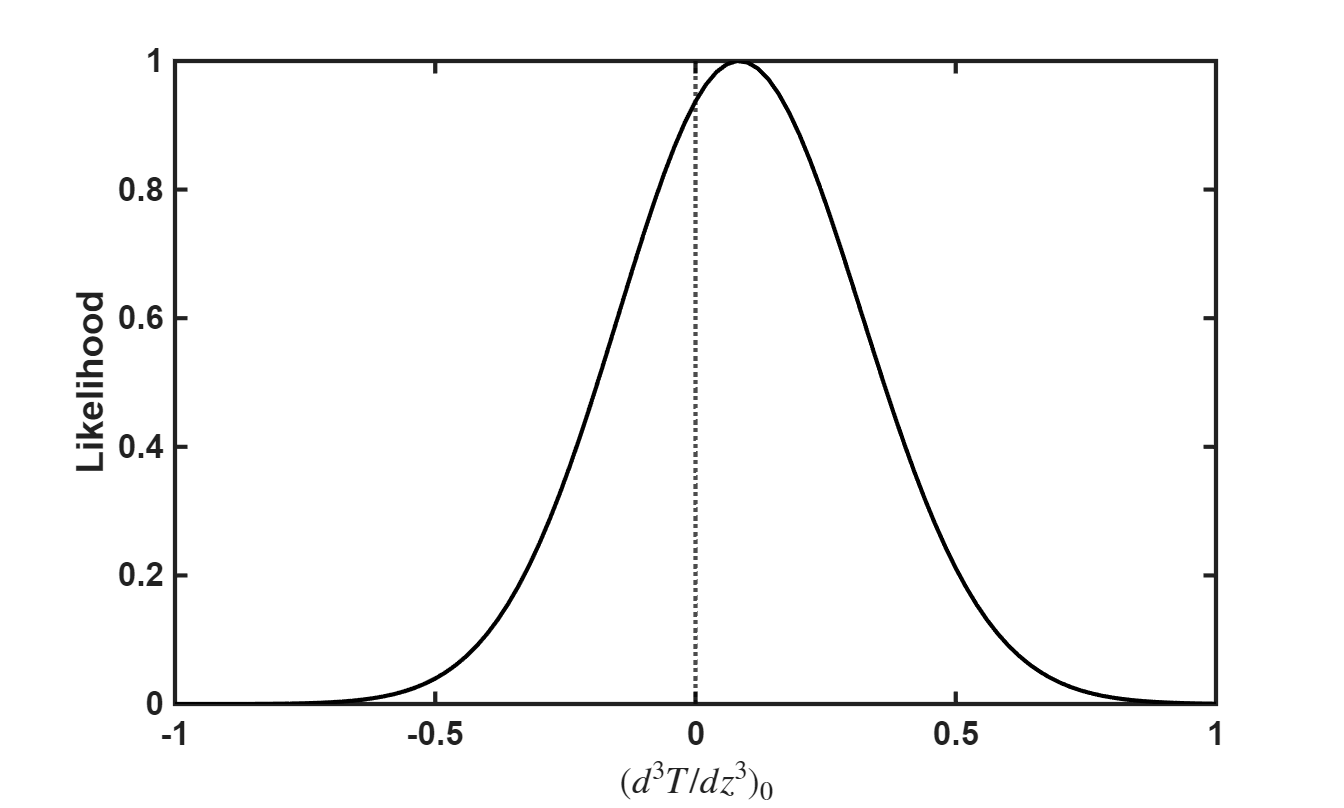}
\includegraphics[width=0.51\columnwidth]{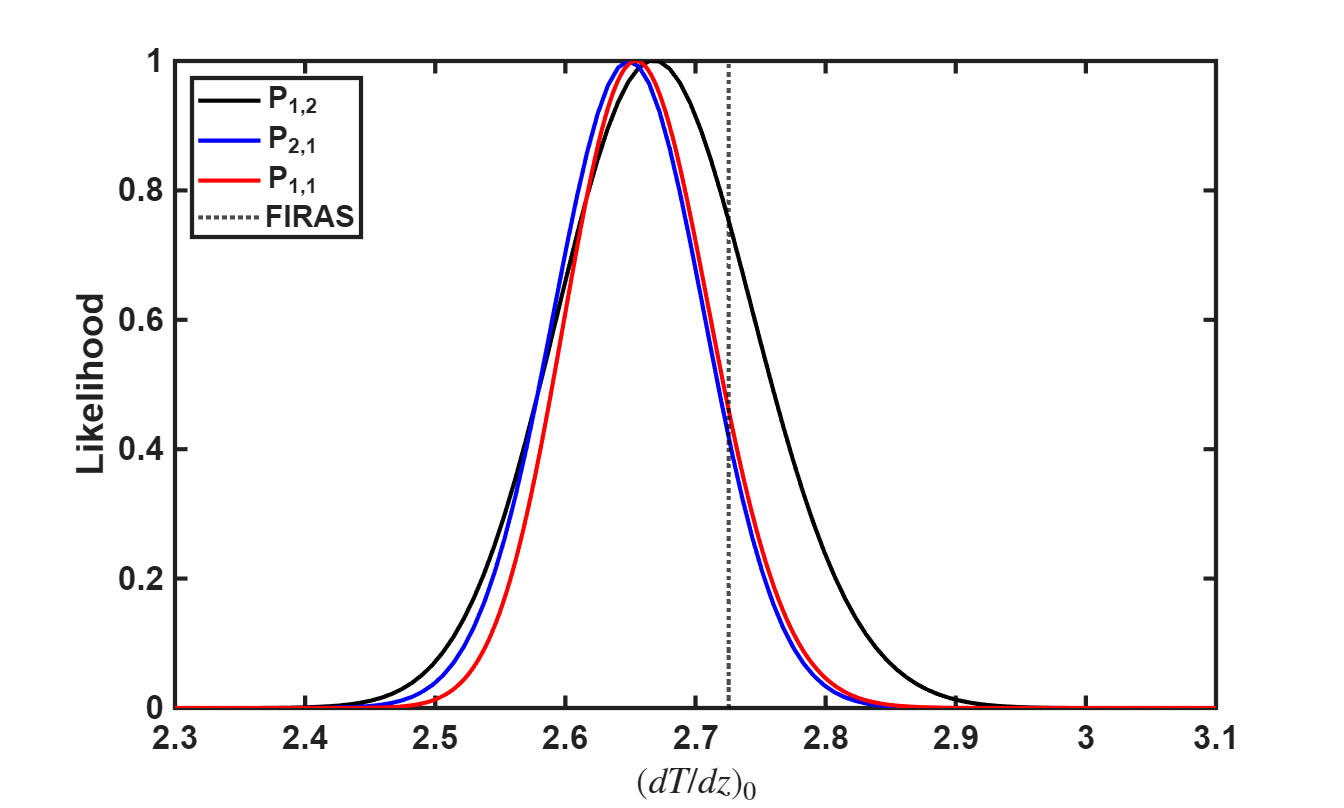}
\includegraphics[width=0.51\columnwidth]{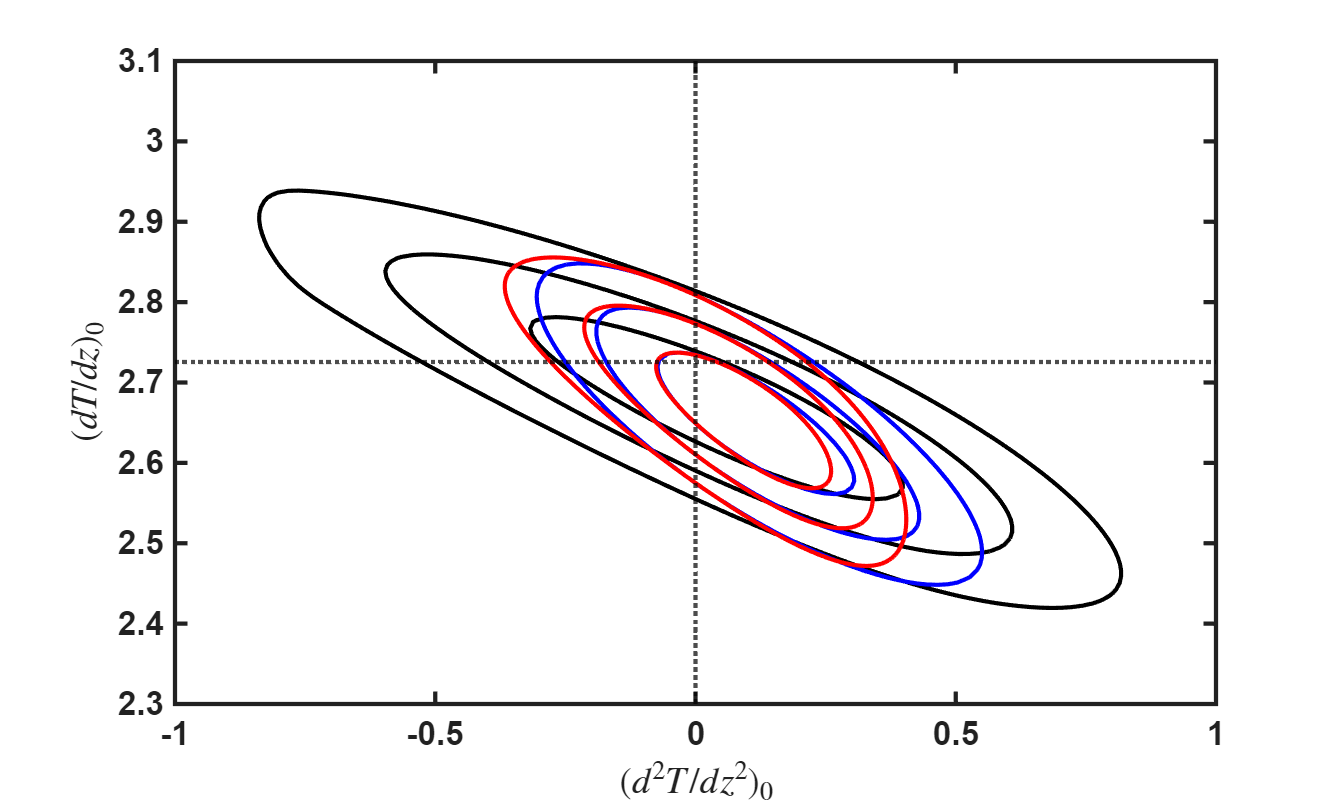}
\includegraphics[width=0.51\columnwidth]{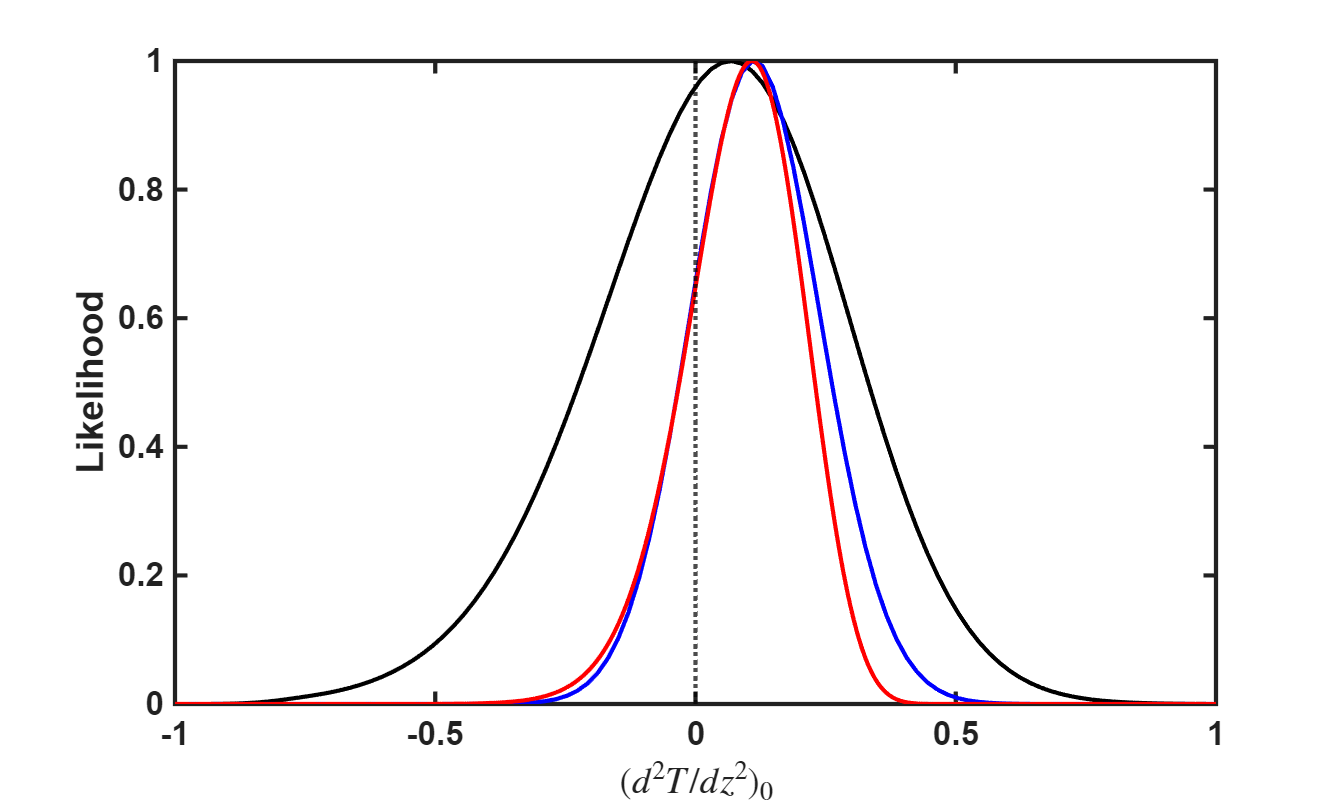}
\includegraphics[width=0.51\columnwidth]{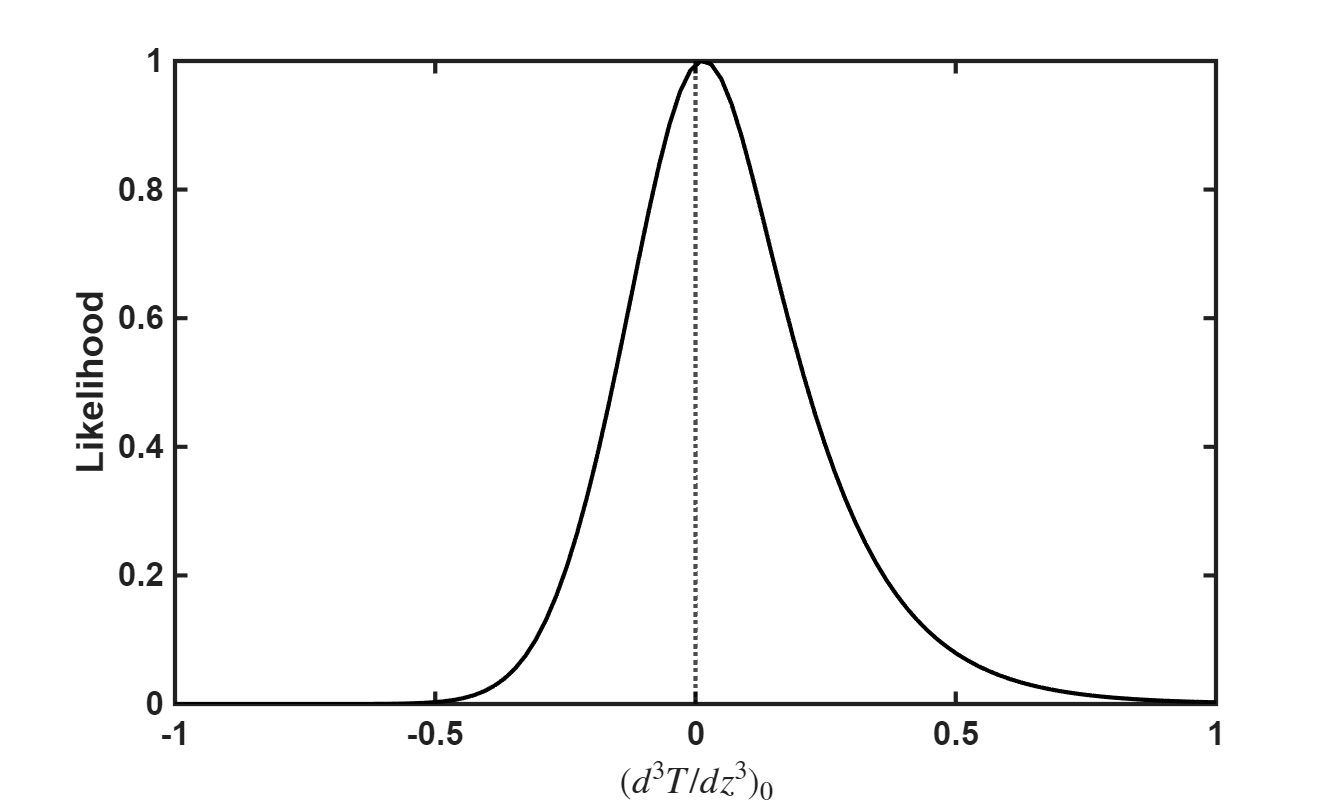}
\caption{Constraints on the derivatives $(d^iT/dz^i)_0$ of Eq. (\ref{series}) for various choices of Taylor series (top panels) and Padé approximants (bottom panels). In the two-dimensional panels, one, two and three sigma confidence levels are shown. Other parameters, when they exist, have been marginalized. The dotted lines correspond to the FIRAS value. Note that the third derivative is not plotted for $P_{2,1}$, cf. the main text.}
\label{figure3}
\end{figure*}

As a preamble, consider a general Taylor expansion of the CMB temperature up to cubic order
\be\label{series}
T_3(z)=T_0+\left(\frac{dT}{dz}\right)_0z+\frac{1}{2}\left(\frac{d^2T}{dz^2}\right)_0z^2+\frac{1}{6}\left(\frac{d^3T}{dz^3}\right)_0z^3+{\cal O}(z^4)\,;
\ee
in what follows we use $T_i$ to denote this series, truncated to order i. The standard law has $T_0=(dT/dz)_0$ and vanishing higher-order coefficients. Abusing notation, we can think of these as $T_i=P_{i,0}$, and it is illustrative to check how each of these is constrained by the present data. The top parts of Table \ref{table3} and Figure \ref{figure3} show these results. Again there is no statistical preference for deviations from the standard law, and a a single additional parameter suffices.

Now, what Pad\'e approximants are suitable? The simplest one, corresponding to a linear order expansion, is
\be
P_{0,1}(z)=\frac{T_0^2}{T_0-(dT/dz)_0z}\,,
\ee
and is evidently unphysical, given that observationally $T_0\sim(dT/dz)_0$. Indeed, all approximants of the form $P_{0,m}$ will have analogous terms in the denominator, and are thus unfit for purpose. Therefore, moving to quadratic order, the only remaining option is
\be\label{pad11}
P_{1,1}(z)=T_0+\frac{2(dT/dz)^2_0z}{2(dT/dz)_0-(d^2T/dz^2)_0z}\,,
\ee
while to cubic order there are two possibilities
\be\label{pad21}
P_{2,1}(z)=T_0+\left(\frac{dT}{dz}\right)_0z+\frac{3(d^2T/dz^2)^2_0z^2}{6(d^2T/dz^2)_0-2(d^3T/dz^3)_0z}
\ee
\be\label{pad12}
P_{1,2}(z)=\frac{(T_0+[(dT/dz)_0+T_0\theta(z)]z)T_0}{T_0+[T_0z-(dT/dz)_0z^2]\theta(z)-1/2(d^2T/dz^2)_0z^2}\,,
\ee
where for simplicity we have defined
\be
\theta(z)=\frac{(d^3T/dz^3)_0T_0-3(dT/dz)_0(d^2T/dz^2)_0}{6(dT/dz)_0^2-3(d^2T/dz^2)_0T_0}\,.
\ee
The former is numerically problematic, since the final term behaves as $0/0$ in the standard model limit. Nominally this can be avoided by a judicious (correlated) choice of priors for the second and third derivatives, e.g. treating the second derivative as a physical parameter to be constrained by the data while the third derivative is a nuisance parameter to be marginalized. In any case, the results will depend on this choice of priors. As for the latter, while cumbersome and therefore not easy to physically intuit, it is mathematically acceptable.

%%%%%%%%%%%%%%%%%%%%
\begin{table}
\centering
\caption{One-sigma constraints on the coefficients of Eq. (\ref{series}), for various choices of Taylor series and Pad\'e approximant. We do not explicitly report the obtained $T_0$ constraint since it always coincides with the FIRAS result.}
\label{table3}
\begin{tabular}{c c c c}
\hline
Function & $(dT/dz)_0$ & $(d^2T/dz^2)_0$ & $(d^3T/dz^3)_0$ \\
\hline
$T_1(z)$ & $2.69\pm0.04$ & - & - \\
$T_2(z)$ & $2.65\pm0.06$ & $0.11\pm0.13$ & - \\
$T_3(z)$ & $2.67\pm0.08$ & $0.03\pm0.27$ & $0.08\pm0.23$ \\
\hline
$P_{1,1}(z)$ & $2.65\pm0.06$ & $0.11^{+0.10}_{-0.12}$ & - \\
$P_{2,1}(z)$ & $2.65\pm0.06$ & $0.11^{+0.13}_{-0.12}$ & Unconstrained \\
$P_{1,2}(z)$ & $2.66\pm0.08$ & $0.07^{+0.22}_{-0.25}$ & $0.01^{+0.16}_{-0.14}$ \\
\hline
\end{tabular}
\end{table}
%%%%%%%%%%%%%%%%%%%%
%%%%%%%%%%%%%%%%%%%%

The bottom parts of Table \ref{table3} and Figure \ref{figure3} show the analogous constraints for the potentially viable Padé approximants. For $P_{2,1}$ we have used the narrow prior $(d^3T/dz^3)_0\in[-10^{-3},+10^{-3}]$, for the aforementioned reasons, and therefore no physically meaningful constraint is obtained for this parameter.

%%%%%%%%%%%%%%%%%%%%%%%%%%%%%%%%%%%%%%%%%%%%%%%%%%%%%%%%%%%%%%%
\section{Removing FIRAS from the analysis}
\label{nofiras}

It is worthy of note that in Table \ref{table3} the slope  $(dT/dz)_0$, although statistically compatible with the FIRAS value, is always smaller than it. Moreover, in Table \ref{table1} the first derivative coefficient posteriors, although compatible with zero, are always skewed to negative values. This raises the question of the impact of FIRAS in these results, and in this section we repeat this analysis without the FIRAS measurement. For simplicity we restrict ourselves to the cases of $T_2(z)$ and $P_{1,1}(z)$, but will separately consider the constraints from Sunyaev-Zeldovich and spectroscopic measurements, as well as their combination.

 %%%%%%%%%%%%%%%%%%%%
\begin{table}
\centering
\caption{One-sigma constraints on the coefficients of Eq. (\ref{series}), for $T_2(z)$ and $P_{1,1}(z)$, without the FIRAS measurement and separatrly considering the Sunyaev-Zeldovich and spectroscopic measurements.}
\label{table4}
\begin{tabular}{c c c c c}
\hline
Function & Data & $T_0$ & $(dT/dz)_0$ & $(d^2T/dz^2)_0$ \\
\hline
{} & SZ & $2.75\pm0.02$ & $2.48\pm0.19$ & $0.61_{-0.56}^{+1.06}$ \\
$T_2(z)$ & Spec & $2.92\pm0.57$ & $2.32\pm0.86$ & $0.32_{-0.50}^{+0.47}$ \\
{} & Joint & $2.74\pm0.02$ & $2.59\pm0.09$ & $0.17\pm0.15$ \\
\hline
{} & SZ & $2.74\pm0.02$ & $2.53_{-0.16}^{+0.18}$ & $0.48^{+0.37}_{-0.47}$ \\
$P_{1,1}(z)$ & Spec &  $2.50_{-0.90}^{+0.48}$ & $2.80_{-0.58}^{+1.26}$ & $0.23^{+0.12}_{-0.43}$ \\
{} & Joint &  $2.74\pm0.02$ & $2.61_{-0.08}^{+0.09}$ & $0.15^{+0.11}_{-0.13}$ \\
\hline
\end{tabular}
\end{table}
%%%%%%%%%%%%%%%%%%%%
\begin{figure*}
\includegraphics[width=0.51\columnwidth]{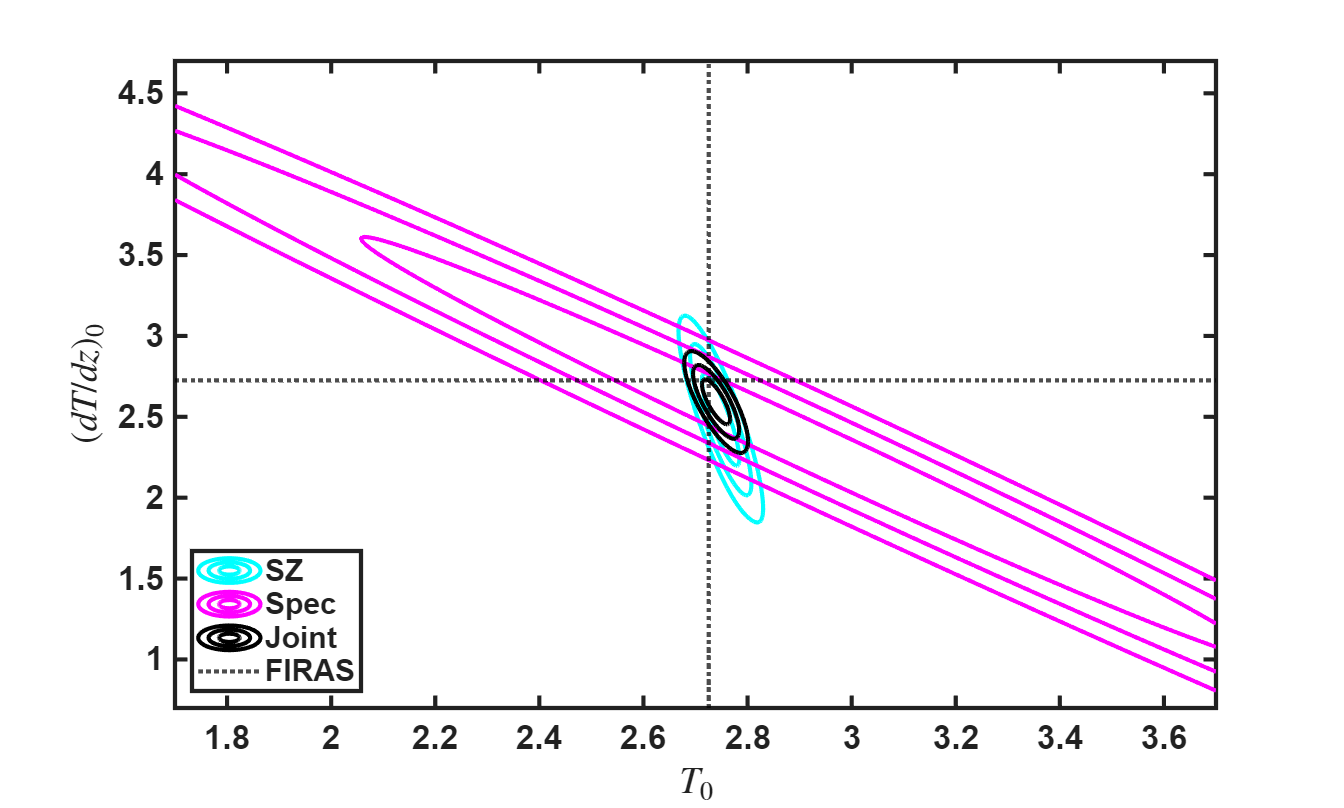}
\includegraphics[width=0.51\columnwidth]{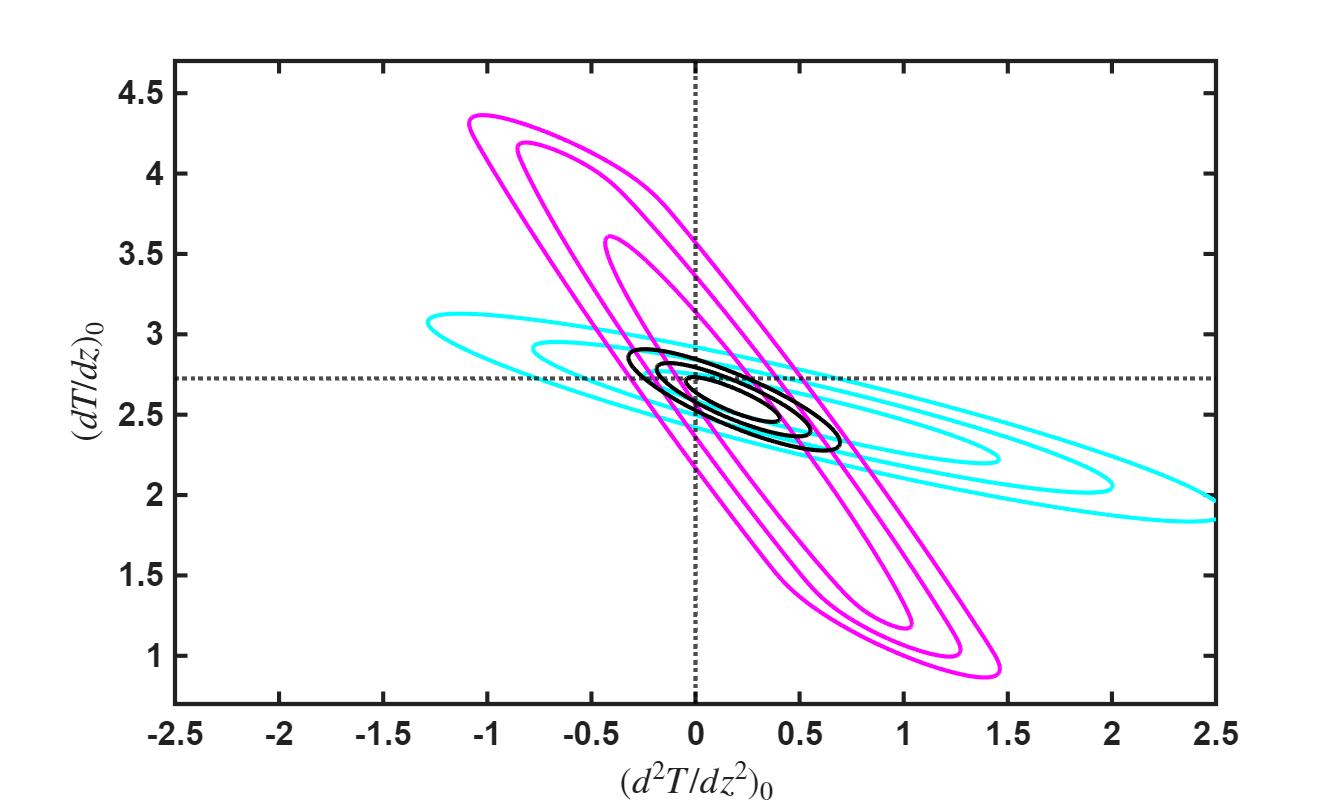}
\includegraphics[width=0.51\columnwidth]{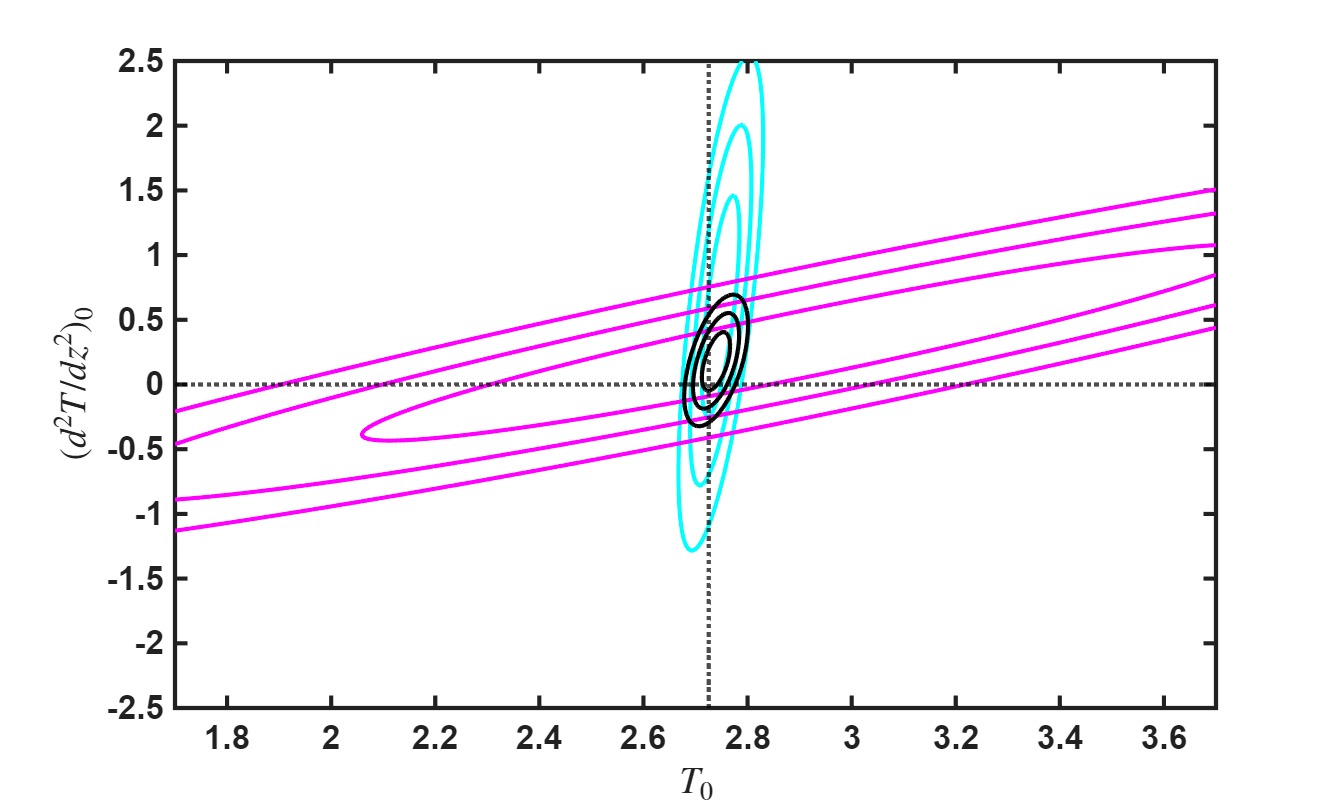}
\includegraphics[width=0.51\columnwidth]{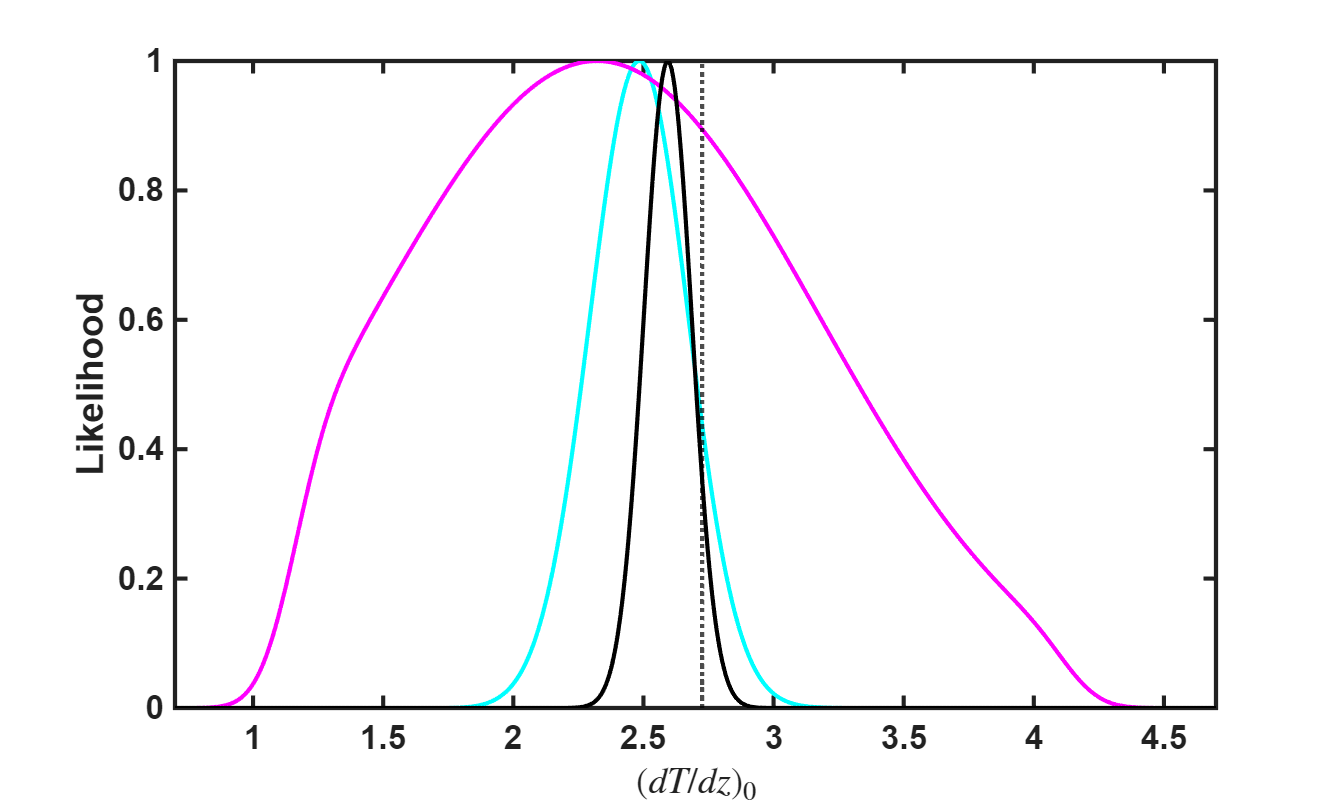}
\includegraphics[width=0.51\columnwidth]{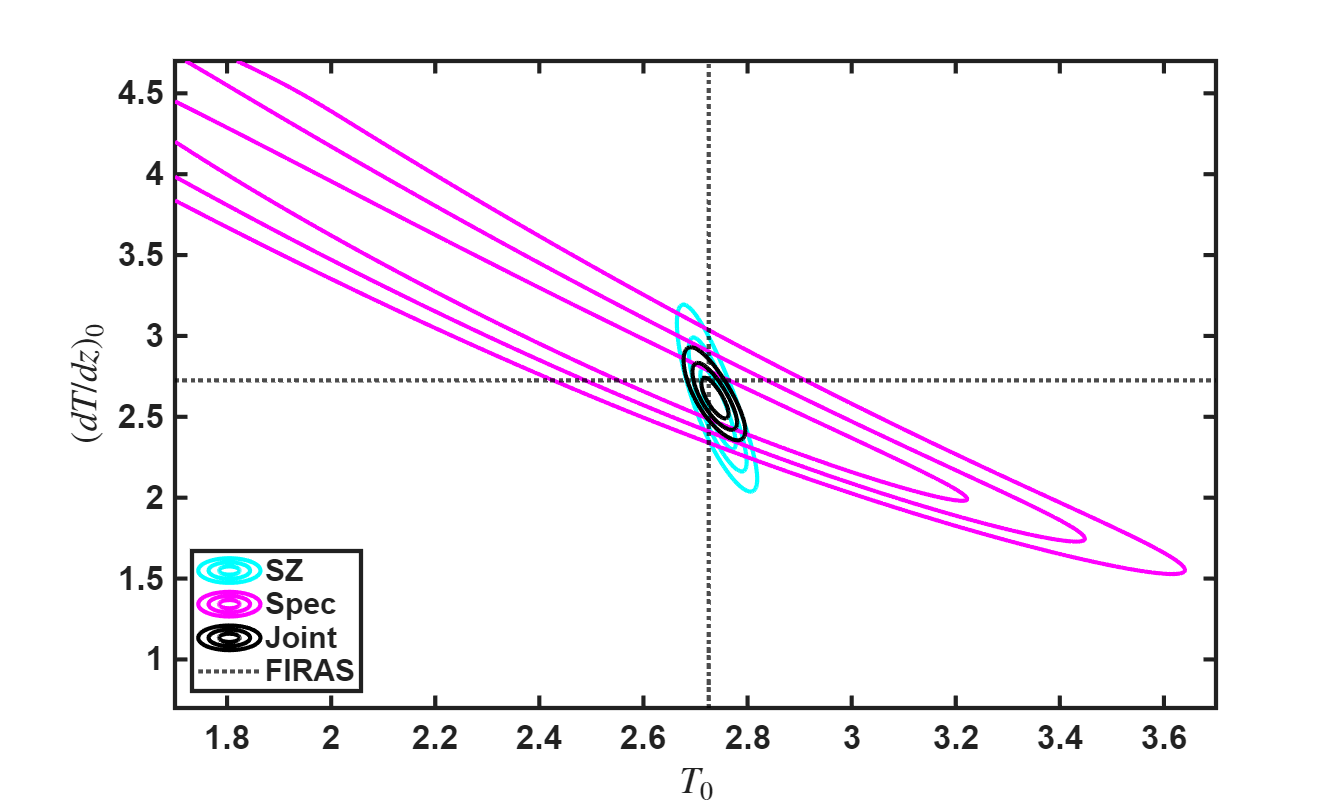}
\includegraphics[width=0.51\columnwidth]{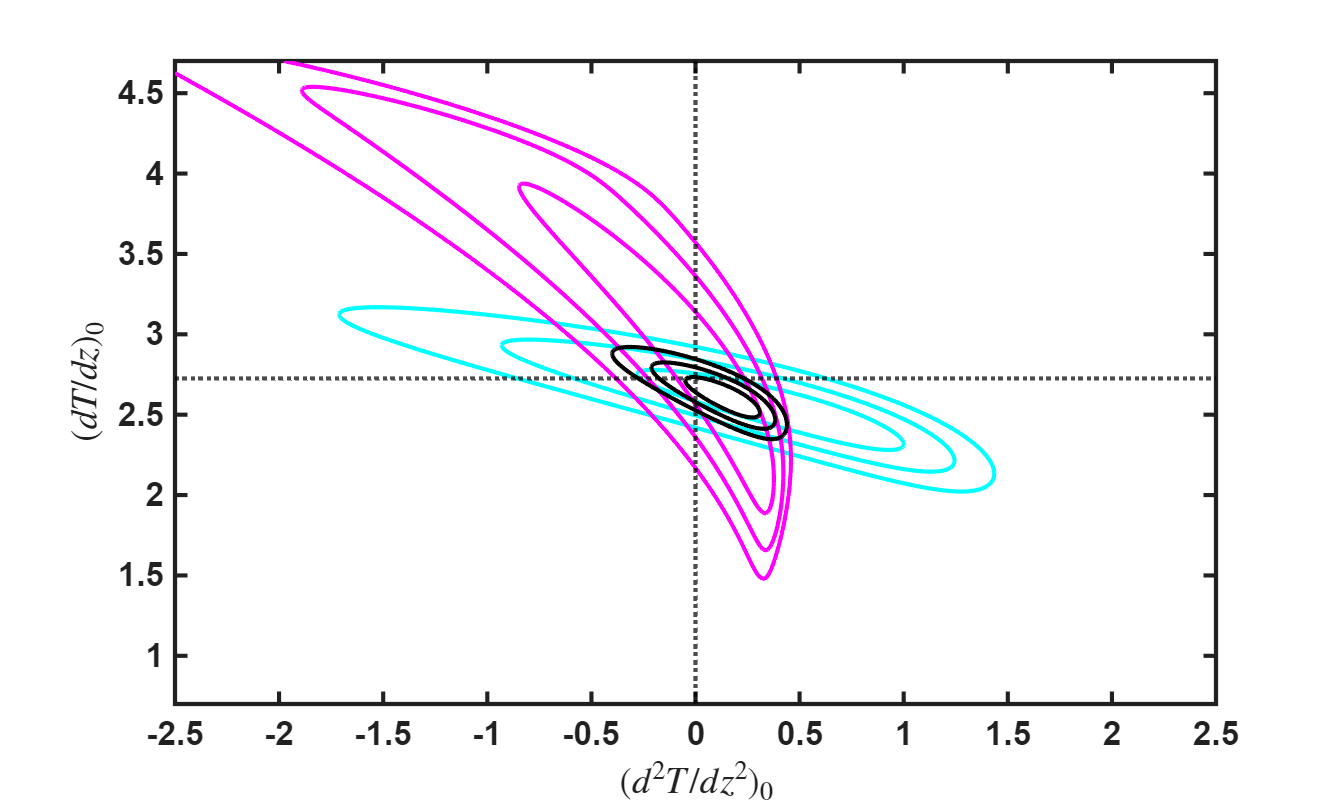}
\includegraphics[width=0.51\columnwidth]{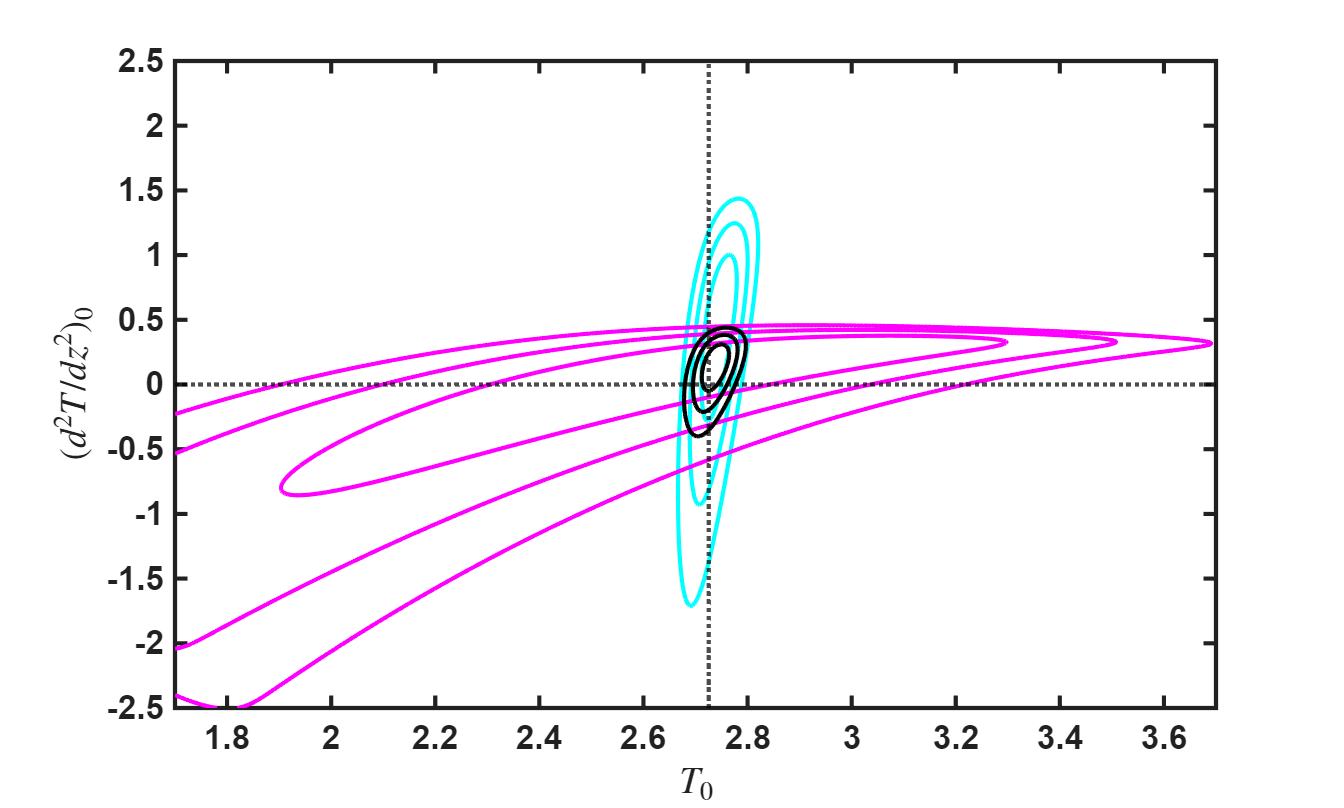}
\includegraphics[width=0.51\columnwidth]{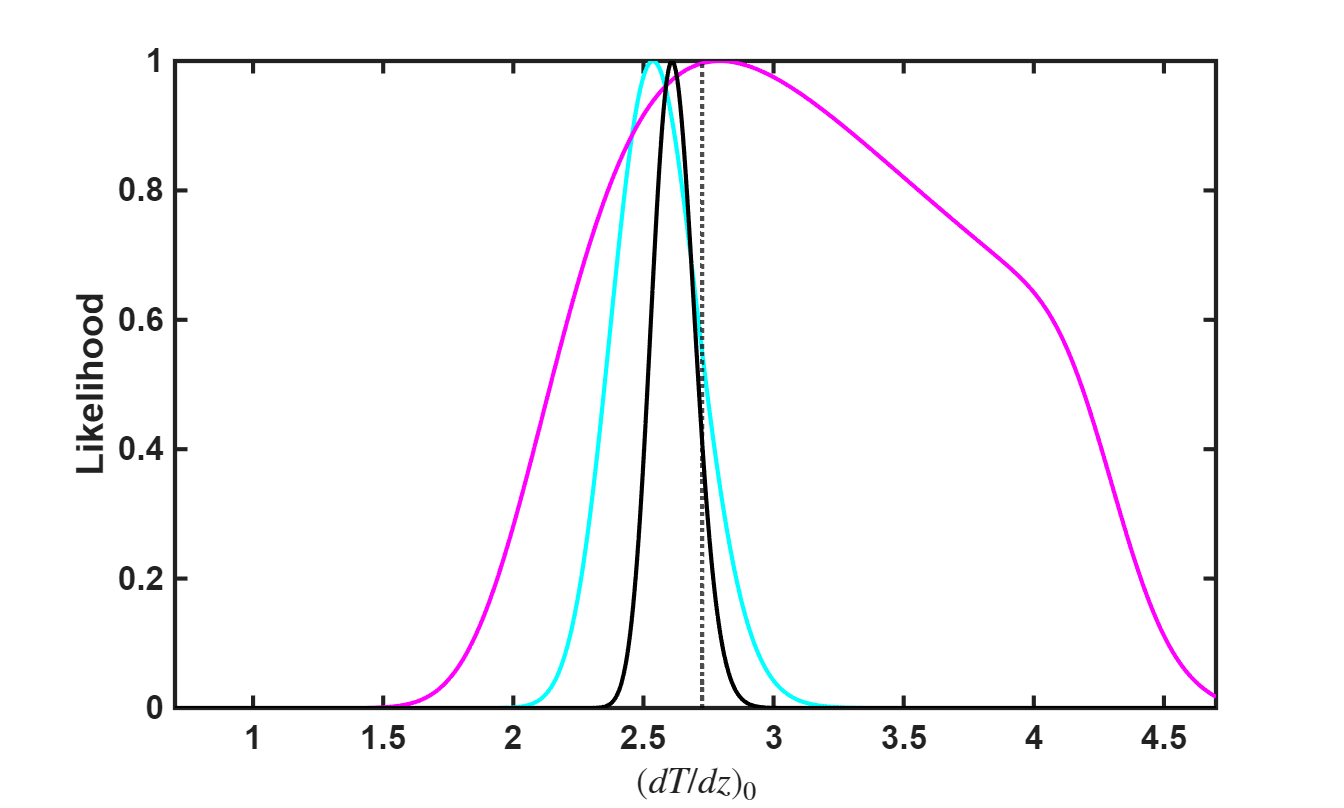}
\caption{Constraints on $T_0$ and its derivatives for $T_2(z)$ (top panels) and $P{1,1}(z)$ (bottom panels), without including the FIRAS value, and considering separately the Sunyaev-Zeldovich and spectroscopic data, as well as the combination of the the two. In the two-dimensional panels, one, two and three sigma confidence levels are shown. Other parameters have been marginalized, and the dotted lines correspond to the FIRAS value.}
\label{figure4}
\end{figure*}
%%%%%%%%%%%%%%%%%%%%

Table \ref{table4} and Figure \ref{figure4} summarize these results. Again there are no statistically significant deviations from the standard behaviour, but an overall preference for a value of $(dT/dz)_0$ which is slightly smaller than the FIRAS one persists. As expected given the (anti)correlations between the parameters, the second derivative $(d^2T/dz^2)_0$ is correspondingly skewed towards slightly positive values, and similarly $T_0$ is skewed towards larger values than the FIRAS one.

This analysis also shows that the Sunyaev-Zeldovich data is significantly more constraining than the high-resolution spectroscopic data. The main  reason is that the latter subset only probes relatively high redshifts; the lack of low-redshift data is a bottleneck for an expansion around redshift zero, since it implies enhanced degeneracies among the model parameters. Nevertheless, the combination of the two subsets is highly desirable, since it leads to much improved constraints on the higher-order parameters. An analogous situation has been quantified by \citet{Rocha} in the context of redshift drift cosmography, where the SKA and the ELT are expected to provide measurements at low and high redshift respectively.

%%%%%%%%%%%%%%%%%%%%%%%%%%%%%%%%%%%%%%%%%%%%%%%%%%%%%%%%%%%%%%%
\section{Conclusions}
\label{concl}

We have applied a range of cosmographic tools to assess the observational status of the temperature-redshift relation. This is a simple but fundamental consistency test of the standard cosmological model, since it is violated in many physically motivated extensions of it---e.g., those featuring photon number non-conservation. Our analysis relied on a range of 63 publicly available measurements of the CMB temperature, spanning the redshift range $0\le z\le6.34$, including the extremely precise FIRAS one \citep{Fixsen} and Sunyaev-Zeldovich and high-resolution spectroscopic measurements.

Our first result is that there is no statistically significant evidence for departures from the standard law, and any such deviations are constrained to percent level---a result which is robust to a range of choices of cosmographic approach, including on the expansion variable and truncation order. This is also the case for the phenomenological---but by now canonical---adiabatic extension model of \citet{Lima1}. Overall, our analysis shows that a single non-standard parameter (which moreover is constrained to be small) suffices to describe current constraints on violations of the standard law. 

We have also shown that some of the traditional cosmography lore, concerning the usefulness of the rescaled redshift $y=z/(1+z)$ or of Pad\'e approximants, does not apply to the present context. The main reason for this is that in traditional cosmography, based on expansions of the scale factor or quantities directly derived from it, all expansion coefficients (viz. the Hubble constant, deceleration parameter, jerk, and so on) are non-zero for most plausible cosmological models. Instead, in our case, there is a finite (and indeed quite small) number of non-zero coefficients, at least in the standard cosmological paradigm, with any additional non-zero coefficients being evidence of new physics. Thus the precision versus accuracy trade-off, which crucially underlies cosmographic approaches, is significantly different in the two cases. We expect that this is also the case for varying fundamental constants cosmography \citep{Cosmography2}.

An intriguing result is that the directly fitted slope of the temperature-redshift relation, $(dT/dz)_0$, although compatible with the one inferred from FIRAS, is consistently skewed towards smaller values. This may have wider cosmological implications. \textit{Inter alia}, a geometric degeneracy between $T_0$ and the Hubble constant ($H_0$) is noted in \citet{Ivanov}, who show that removing the FIRAS $T_0$ prior makes Planck and local $H_0$ measurements less discrepant. Conversely, combining Planck and local $H_0$ data leads to a $T_0$ measurement with a 3 sigma tension with FIRAS. This is also superficially noted in \citet{ACT2}, and worthy of further exploration.

Our results are dominated by the lower-redshift SZ data, Forthcoming facilities such as ELT-ANDES, can help probe higher redshifts \citep{ANDES}. Another standard model cornerstone, the distance duality relation, is intimately related with the temperature-redshift relation: violations of the two often occur together and are not independent \citep{Duality}. Measurements of the CMB temperature can therefore be added to traditional distance-duality relation probes, leading to more stringent constraints---either for specific models of interest, or in a generalized cosmographic approach.

%%%%%%%%%%%%%%%%%%%%%%%%%%%%%%%%%%%%%%%%%%%%%%%%%%

\section*{Acknowledgements}

Discussions on the topic of this work with Ana Isabel Martins and Elia Stefano Battistelli during the cosmology session of the 2026 Rencontres de Moriond are gratefully acknowledged.

This work was financed by Portuguese funds through FCT (Funda\c c\~ao para a Ci\^encia e a Tecnologia) in the framework of the project 2022.04048.PTDC (Phi in the Sky, DOI 10.54499/2022.04048.PTDC). CJM also acknowledges FCT and POCH/FSE (EC) support through Investigador FCT Contract 2021.01214.CEECIND/CP1658/CT0001 (DOI 10.54499/2021.01214.CEECIND/CP1658/CT0001).

\section*{Data Availability}

The data analysed in this work is publicly available in the cited references.

\bibliographystyle{mnras}
\bibliography{cmb} % if your bibtex file is called example.bib
%%%%%%%%%%%%%%%%%%%%%%%%%%%%%%%%%%%%%%%%%%%%%%%%%%
\bsp	% typesetting comment
\label{lastpage}
\end{document}